\pdfoutput=1

\documentclass[conference,9pt]{IEEEtran}

\usepackage{algpseudocode}
\algtext*{EndWhile}
\algtext*{EndIf}
\algtext*{EndFor}
\algtext*{EndProcedure}
\usepackage{amsmath}
\usepackage{amssymb}
\usepackage{array}
\usepackage{algorithm}
\usepackage{bm}
\usepackage[bookmarks=false, hidelinks]{hyperref}
\usepackage[noadjust]{cite}
\usepackage[font=footnotesize, bf]{caption}
\usepackage{color}
\usepackage{comment}
\usepackage{dblfloatfix}
\usepackage[inline]{enumitem}
\usepackage{epsfig}
\usepackage{etoolbox}
\usepackage{fancybox}
\usepackage{float}
\usepackage{fixltx2e}
\usepackage{graphicx}
\usepackage{listings}
\usepackage{mathrsfs}
\usepackage{multirow}
\usepackage{multicol}
\usepackage{pifont}
\usepackage{placeins}
\usepackage{rotating}
\usepackage{setspace}
\usepackage{soul}
\usepackage{threeparttable}
\usepackage{times}
\usepackage{url}
\usepackage{upgreek}
\usepackage{verbatim}
\usepackage{wrapfig}
\usepackage[usenames,dvipsnames]{xcolor}
\usepackage{authblk}
\usepackage[]{siunitx}
\usepackage[super]{nth}
\usepackage{tabularx}
\usepackage{supertabular}
\usepackage{tikz}
\usepackage{bbm}
\usepackage{blindtext}
\usepackage{hhline} 
\usepackage[export]{adjustbox}
\usepackage{cleveref}

\definecolor{lightgreen}{RGB}{198, 224, 183}
\definecolor{lightorange}{RGB}{242, 178, 136}

\graphicspath{{./_fig/}}

\makeatletter
\renewcommand\footnoterule{%
  \kern-3\p@
  \hrule\@width0.4\columnwidth
  \kern2.6\p@}
  \makeatother

\newcommand{\todo}[1]{\footnote{\textbf{\color{red}{TODO:}} #1}}

\usepackage[hang,flushmargin]{footmisc}
\usepackage{bbding}
\usepackage{lettrine}
\usepackage{xcolor,colortbl}

\definecolor{cyan}{RGB}{213,229,255}
\definecolor{orangeShallow}{RGB}{255,190,0}

\usepackage{subfig}
\usepackage{tikz}
\usepackage{xcolor}
\newcommand*\circled[1]{\tikz[baseline=(char.base)]{
            \node[shape=circle,fill,inner sep=1pt,scale=0.8] (char) {\textcolor{white}{#1}};}}

\begin{document}

%\title{RTLer: Predicting Post-Synthesis PPA for Any Given RTL Design\vspace{-.2in}}

%Prestimator: A \underline{Pre}-Synthesis PPA Estimation Framework for Any RTL Design?
%PRestimator: A \underline{P}re-Synthesis \underline{R}TL PPA Estimation Framework?
%Or maybe Restimator (R for RTL)?

%\title{RTLer: A Pre-Synthesis PPA Estimation Framework for Any RTL Design\vspace{-.2in}}

%\title{PeekRTL: A \underline{P}re-Synth\underline{e}sis PPA \underline{E}stimation Framewor\underline{k} for Any \underline{RTL} Design\vspace{-.2in}}

\title{MasterRTL: A Pre-Synthesis PPA Estimation Framework\\for Any RTL Design\vspace{-.11in}}

%\title{MasterRTL: Maximizing Design Efficiency with a PPA Estima\underline{t}ion Fram\underline{e}wo\underline{r}k for any RTL Design\vspace{-.2in}}

%\title{MasterRTL: \underline{Ma}ximizing Design Efficiency by Estimating the PPA of any \underline{RTL} Design\vspace{-.2in}}

%\title{MasterRTL: I\underline{m}proving Design Efficiency with \underline{A} PPA E\underline{s}tima\underline{t}ion Fram\underline{e}work fo\underline{r} any \underline{RTL} Design\vspace{-.2in}}

% Startle

\author[]{ \fontsize{11}{11}\selectfont Wenji Fang$^{1,2}$, Yao Lu$^2$, Shang Liu$^2$, Qijun Zhang$^2$,  Ceyu Xu$^3$, Lisa Wu Wills$^3$, Hongce Zhang$^{1,2}$\textsuperscript{*}, Zhiyao Xie$^2$\textsuperscript{*}\vspace{-5pt}}

\affil[]{\fontsize{10}{10}\selectfont $^1$Hong Kong University of Science and Technology (Guangzhou), \\$^2$Hong Kong University of Science and Technology, $^3$Duke University\vspace{-6pt}}

\affil[]{$\textsuperscript{*}$Corresponding Author: \{hongcezh, eezhiyao\}@ust.hk\vspace{-12pt}}

\maketitle
%\begingroup\renewcommand\thefootnote{*}
%\footnotetext{Corresponding Author}
%\endgroup

\begin{spacing}{1.0}
\begin{abstract}

In modern VLSI design flow, the register-transfer level (RTL) stage is a critical point, where designers define precise design behavior with hardware description languages (HDLs) like Verilog. Since the RTL design is in the format of HDL code, the standard way to evaluate its quality requires time-consuming subsequent synthesis steps with EDA tools. This time-consuming process significantly impedes design optimization at the early RTL stage. Despite the emergence of some recent ML-based solutions, they fail to maintain high accuracy for any given RTL design. 
% HZ: Register-transfer level (RTL) is a critical design stage in modern digital VLSI flow, where designers precisely define microarchitecture using hardware description languages (HDLs), such as Verilog. Evaluating RTL design quality usually requires going through the tedious logic synthesis step with EDA tools, which is often too time-consuming for rapid design-space exploration and optimization at the early RTL stage. Meanwhile, prior works on ML-based design quality evaluation failed to deliver a high prediction accuracy across arbitrary RTL designs.
In this work, we propose an innovative pre-synthesis PPA estimation framework named MasterRTL. It first converts the HDL code to a new bit-level design representation named the simple operator graph (SOG). By only adopting single-bit simple operators, this SOG proves to be a general representation that unifies different design types and styles. The SOG is also more similar to the target gate-level netlist, reducing the gap between RTL representation and netlist. In addition to the new SOG representation, MasterRTL proposes new ML methods for the RTL-stage modeling of timing, power, and area separately. Compared with state-of-the-art solutions, the experiment on a comprehensive dataset with 90 different designs shows accuracy improvement by 0.33, 0.22, and 0.15 in correlation for total negative slack (TNS), worst negative slack (WNS), and power, respectively. 
\end{abstract}
\end{spacing}
%Some most recent works convert design RTL to . 
%Most recently, some methods are proposed to process . 
%highly timing-consuming. 
%Sufficient design optimization at the RTL stage is highly desired but extremely difficult, due to the difficulty in evaluating RTL design.  

%The timing model is the first... =
%It applies to unknown new RTL designs . 
%It is desired to sufficiently optimize at this stage, before entering subsequent . 
% design optimization at an earlier stage is always desired. T
%designers spend a great effort in optimizing their at the register-transfer level (RTL) stage. 

\iffalse
\todo{1. representation (Word vs Bit vs Netlist)\

      2. graph propagation algorithm\
      
      3. timing path-level rule for different design scale\
      
      4. Ablation Study
      
      5. trade-off
      
      6. vector-based power}
\fi

\section{Introduction}

% In the modern VLSI design flow,

%The design and optimization of very large-scale ASIC is an immensely complex process, involving a huge space of design options. After making all design decisions, design engineers generate detailed hardware design descriptions using hardware description languages (HDL) like Verilog or VHDL. 

%In the modern VLSI design flow, the register-transfer level (RTL) stage is a key point, where designers spend great efforts in writing precise design behavior descriptions using hardware description languages (HDL) like Verilog, VHDL, or even Chisel~\cite{bachrach2012chisel}. At such an early stage, design engineers face a huge space of design options with the maximum flexibility to make almost any legal fine-grained design decisions, which will determine the ultimate ASIC design quality in terms of power, powerformance, and area (PPA). Ideally, designers should sufficiently optimize their RTL design at this stage, since it is extremely difficult to remedy low-quality RTL in subsequent downstream synthesis stages. 

% \yao{I will revise this Intro later today}

In modern VLSI design flows, the register-transfer level (RTL) stage is a critical point, where designers devote significant effort to crafting precise design behavior descriptions using hardware description languages (HDLs) such as Verilog, VHDL, and Chisel~\cite{bachrach2012chisel}. At this early stage, design engineers face a vast design space with maximum flexibility, allowing them to make virtually any fine-grained design decisions that will affect the ultimate quality of the ASIC design in terms of power, performance, and area (PPA). Ideally, designers should optimize their RTL designs sufficiently at this stage, since it is extremely challenging, if not impossible, to remedy low-quality RTL in downstream synthesis stages.

%Then the design RTL is implemented by going through the time-consuming subsequent logic synthesis and layout stages, relying on full-fledged commercial electronic design automation (EDA) tools. 

%However, since an RTL design is in the format of HDL code, its ultimate design quality cannot be directly evaluated in a short time. In a standard VLSI design flow, designers have to go through the time-consuming subsequent synthesis stages, relying on full-fledged commercial electronic design automation (EDA) tools. %As a result, designers need to invoke time-consuming synthesis tools frequently to evaluate their design RTL. Based on the evaluation results, designers may optimize the RTL, then they have to go through the time-consuming evaluation process again. Such iteration continues until the optimization finishes. Such inefficiency in the RTL design process obviously hinders the optimization of design quality and prolongs the total turnaround time. 

Despite the critical importance of optimizing RTL designs, it is very difficult to evaluate the RTL design quality, considering an RTL design is still in the format of HDL code. In a standard VLSI design flow, designers have to go through the time-consuming subsequent synthesis or even layout stages, relying on full-fledged commercial electronic design automation (EDA) tools to evaluate the design quality based on netlists or layouts. For complex industrial designs, the logic synthesis could take more than one day and the layout process can easily further take several days. To make things worse, designers often need to frequently invoke synthesis tools to implement and evaluate their RTL designs, optimize the RTL code based on the results, and then repeat the evaluation process until optimization is complete. This iterative process significantly prolongs the total turnaround time and hinders the optimization of design quality, making the RTL design process extremely inefficient.

% Designers must rely on full-fledged commercial electronic design automation (EDA) tools, which can be time-consuming to use. Designers may need to frequently invoke synthesis tools to evaluate their RTL design, optimize it based on the results, and then repeat the evaluation process until optimization is complete. This iterative process can prolong the total turnaround time and hinder the optimization of design quality, making the RTL design process inefficient.

%In recent years, many customized machine learning (ML) methods are explored to provide design quality predictions as early design feedbacks~\cite{rapp2021mlcad}. However, most ML methods target prediction on gate-level netlist or layouts, with the earlier RTL stage being less explored~\cite{rapp2021mlcad}. In existing ML for EDA solutions, gate-level netlists are commonly represented as graphs and processed with graph neural networks (GNNs), and layouts are represented as two-dimensional matrices and processed with convolutional neural networks (CNNs). But since the RTL design is in the HDL code format instead of a common data structure, there is no consensus on the best way to handle it yet. Some works~\cite{xie2020fist, liang2021flowtuner} can only tune the design flow of a specific design without examining RTL details. These models have to be retrained for every new design. Similary, most RTL-stage power models~\cite{xie2021apollo, zhou2019primal, kim2019simmani, xie2022deep, yang2015early} cannot be generalized to new designs. 

In recent years, many customized machine learning (ML) methods have been explored to predict design quality as early design feedbacks~\cite{rapp2021mlcad}. However, most ML methods make predictions based on gate-level netlists or layouts, with the earlier RTL stage receiving less attention~\cite{rapp2021mlcad}. Existing ML solutions commonly represent gate-level netlists as graphs and process them with graph neural networks (GNNs), while layouts are represented as two-dimensional matrices and processed with convolutional neural networks (CNNs). However, since the RTL design is in HDL code format instead of common data structures, there is no consensus on the best way to represent and handle such RTL designs yet. Some works~\cite{xie2020fist, liang2021flowtuner} only tune the design flow of specific designs without examining RTL details, therefore requiring retraining the ML model for every new design. Similarly, most RTL-stage power models~\cite{xie2021apollo, zhou2019primal, kim2019simmani, xie2022deep, yang2015early} cannot be generalized to new designs. Other works~\cite{lopera2021rtl,lopera2022applying} develop RTL-stage timing models based on Neural Networks. However, these models only support combinational circuits and are trained using specific design variations generated by an RTL generator. Besides the RTL-stage PPA modeling, which is the focus of this work, there are models targeting even earlier architectural stages~\cite{davis2021fast, zhai2021mcpat, wu2022high, ustun2020accurate, shao2014aladdin, zhang2023panda}. But it is even harder for them to directly generalize to unknown new designs, due to the lack of RTL details. 

%Architectural-level methods~\cite{davis2021fast, zhai2021mcpat} evaluate design qualities without examining RTL implementations, but they are not well generalized either. Other works~\cite{wu2022high, shao2014aladdin, ustun2020accurate} evaluate the design qualities at the high-level synthesis (HLS) stage. However, they only target FPGA platforms~\cite{wu2022high, ustun2020accurate} or accelerators~\cite{shao2014aladdin}. Typically, pre-RTL solutions are harder to generalize to unknown new designs, due to the lack of RTL implementation details. 

%, in the format of FIRRTL~\cite{Li:EECS-2016-9} or RTLIL~\cite{wolf2013yosys}. 
%neither the RTL representation nor the processing methods are sufficiently optimized for 
%  which leave a huge room for accuracy improvement.

% leaving a huge room for accuracy improvement. 

Most recently, a few more general ML methods~\cite{sengupta2022good, xu2022sns} are proposed to predict design qualities at the RTL stage. They first convert design RTL to representations such as the abstract syntax tree (AST), and then evaluate design PPA either based on all register trees~\cite{sengupta2022good} or randomly sampled paths~\cite{xu2022sns} extracted from AST-alike representations. 
However, their accuracy on unknown new RTL designs is still limited for several reasons. 
First, the AST-alike representation adopted in these works~\cite{xu2022sns, sengupta2022good} is simply the initial data format used by traditional synthesis tools. It is not an ideal data format to support ML solutions. As we will demonstrate in Subsection~\ref{sec:representation}, the discrepancies of different design types are obvious in AST-alike representations, limiting the model generalization ability. 
Second, these works~\cite{sengupta2022good, xu2022sns} process the representations with several unreasonable operations.
%Second, there are several unreasonable operations in how they process the representations~\cite{sengupta2022good, xu2022sns}. 
For example, in~\cite{sengupta2022good}, we find notable undesired duplications among register trees when counting the design logic. In~\cite{xu2022sns}, there is a significant gap between the pseudo training paths and real paths extracted from target inference designs. 
%An detailed inspection of these prior works is provided in Subsection~\ref{sec:baseline}.
Subsection~\ref{sec:baseline} provides a detailed inspection of these prior works.
\looseness=-1

In this work, we propose a new RTL-stage PPA modeling framework named MasterRTL, which achieves significantly higher accuracy over prior works~\cite{sengupta2022good, xu2022sns} when applied to new RTL designs. It is the first work that supports the cross-design RTL evaluation on all major PPA qualities, including both total negative slack (TNS) and worst negative slack (WNS), both vector-less and vector-based power analysis results, and the gate area\footnote{In comparison, \cite{sengupta2022good} only evaluates TNS and vector-less power, \cite{xu2022sns} only evaluates WNS, vector-less power, and area. Most other RTL-stage models~\cite{xie2020fist, liang2021flowtuner, xie2021apollo, zhou2019primal, kim2019simmani, xie2022deep, yang2015early} are not cross-design, requiring retraining on new designs.}. %It involves two major steps. First, the RTL code should be converted to a representation for processing. Then ML models should be developed for the RTL-stage PPA modeling task. Specifically, this work 
It primarily answers two key unsolved questions in the RTL-stage PPA modeling problem: %after our detailed exploration: 

\begin{description}
    \item Q1. What is the most appropriate data format of RTL design (i.e. RTL representation) that best supports ML methods? \vspace{.02in}%What is the most ML-friendly representation that should be adopted when processing the RTL format? 
    \item Q2. Based on the RTL representation, how to capture the key patterns to estimate each design objective? %what is the most appropriate ML method to estimate each design objective? %at the RTL stage? %Our observations are summarized below. 
\end{description}

%  in this RTL-stage PPA modeling task
%MasterRTL answers both questions based on the following key ideas. %below. 

% as the basis of all PPA predictions,
For question Q1, since we target a cross-design ML-based method applicable to any RTL design, an appropriate RTL representation should be maximally \emph{similar} to the ultimate gate-level netlist, and be as \emph{general} as possible. Such a \emph{general} representation will unify different design types, thus maximizing the ML model's accuracy on unknown new designs. MasterRTL adopts a new bit-level design representation named the simple operator graph (SOG), with only fundamental single-bit logic operations. Compared with the AST-alike representations in prior works~\cite{xu2022sns, sengupta2022good}, it better unifies different RTL design types and styles and thus enables a higher cross-design model accuracy for almost all design objectives and ML methods.\looseness=-1

For question Q2, since the mechanisms behind ground-truth power, performance, and area measurement are largely different, instead of adopting similar input features for different tasks in prior works~\cite{xu2022sns, sengupta2022good}, we customize different estimation methodologies for timing, power, and area separately. Specifically, among all RTL-stage cross-design methods, our timing model is the first to explicitly capture the critical path and the corresponding delay between any pair of registers. This is enabled by our SOG representation's \emph{consistency} in register mapping with the netlist. Our power model is also the first to integrate toggle rate information as features, thus supporting unified predictions on both vector-based and vector-less power values. It is also the first cross-design RTL power model that employs module-level power evaluation, significantly augmenting the number of power labels for training. \looseness=-1

%For example, the power consumption depends on the total number of toggled capacitances in the design, while TNS/WNS calculation depends on the accumulated delay on all critical paths. Thus, our power modeling method captures toggle rates, while the performance modeling method tries to capture the longest paths among registers. 

%Therefore, instead of adopting similar input features for different tasks in prior works~\cite{xu2022sns, sengupta2022good}, we customize different estimation methodologies for power, performance, and area estimation separately. 

% %As a result, power models should effictively . 
% The gate area measurement is simpler, strongly correlate with the number of gates. 

%Third, . 

%There is no uniform model that 
%maximizing the similarity between different designs. In this way, the ML model can . Second, . 
% the extracted register trees lead to
%However, there can be serious overlaps in captured register trees. 
%However, the exploration just started and there is no consensus on . 

%no commericial EDA tools support direct evaluation of the RTL quality. 
%Existing EDA tools do not support designers to evaluate their RTL quality, unless going through the time-consuming synthesis steps. 
%These subsequent design stages are more automatic while time-consuming.  
%by going through synthesis and layout stages, where the overall automation level is relatively high. 

Our contributions in this work are summarized below:
\begin{itemize}
    \item We propose an open-sourced new framework named MasterRTL to efficiently evaluate all PPA values of any given design RTL\footnote{It is open-sourced in https://github.com/fangwenji/MasterRTL.git}. Evaluated on our comprehensive dataset with 90 different RTL designs, it achieves 0.33, 0.22, and 0.15 higher absolute values in correlation $R$ for TNS, WNS, and power estimations, than state-of-the-art solutions. %The proposed framework is open-source at:~\cite{masterrtl}.
    \item We adopt a new bit-level RTL representation named SOG. Compared to ASTs used in prior works, it is a significantly more general representation that unifies different design types, and reduces the gap to the ultimate gate-level netlist. It is a highly ``ML-friendly'' representation that can be adopted by follow-up ML solutions involving cross-design RTL processing tasks. 
    %all ML solutions should adopt .  
    % leading to better cross-design PPA prediction accuracy. 
    % Compared with word-level representation adopted in prior works, i%proves to correlate significantly better with post-synthesis design PPA values. 
    \item Based on SOG, we customize algorithms for each design objective, capturing their different mechanisms. Among cross-design RTL-stage methods, MasterRTL is the first to capture detailed critical-path information in timing modeling, and the first to integrate toggle rate and module-level information in power modeling.\looseness=-1 %Specifically, we develop the first detailed timing model that tries to capture all the longest paths between registers at the RTL stage. 
%    \item We construct a comprehensive dataset with 90 different RTL designs from 7 different sources. Based on this comprehensive dataset, we perform detailed comparisons with prior works with sufficient ablation studies.
    \item We further study the impact of different logic synthesis options and the extra placement step on the target design PPA. We extend MasterRTL to predict post-placement PPA values. 
    \item Finally, we explore a new data augmentation technique that generates new pseudo RTL designs from scratch. We demonstrate its potential in mitigating the circuit data availability problem.   
    %provide a detailed analysis on the limitation of prior works. %analyze the unreasonable mechanisms adopted in prior works. %The adopts an essentially more reasonable. 
    %\item 
\end{itemize}

\begin{figure}[!t]
  \centering
  \includegraphics[width=1\linewidth]{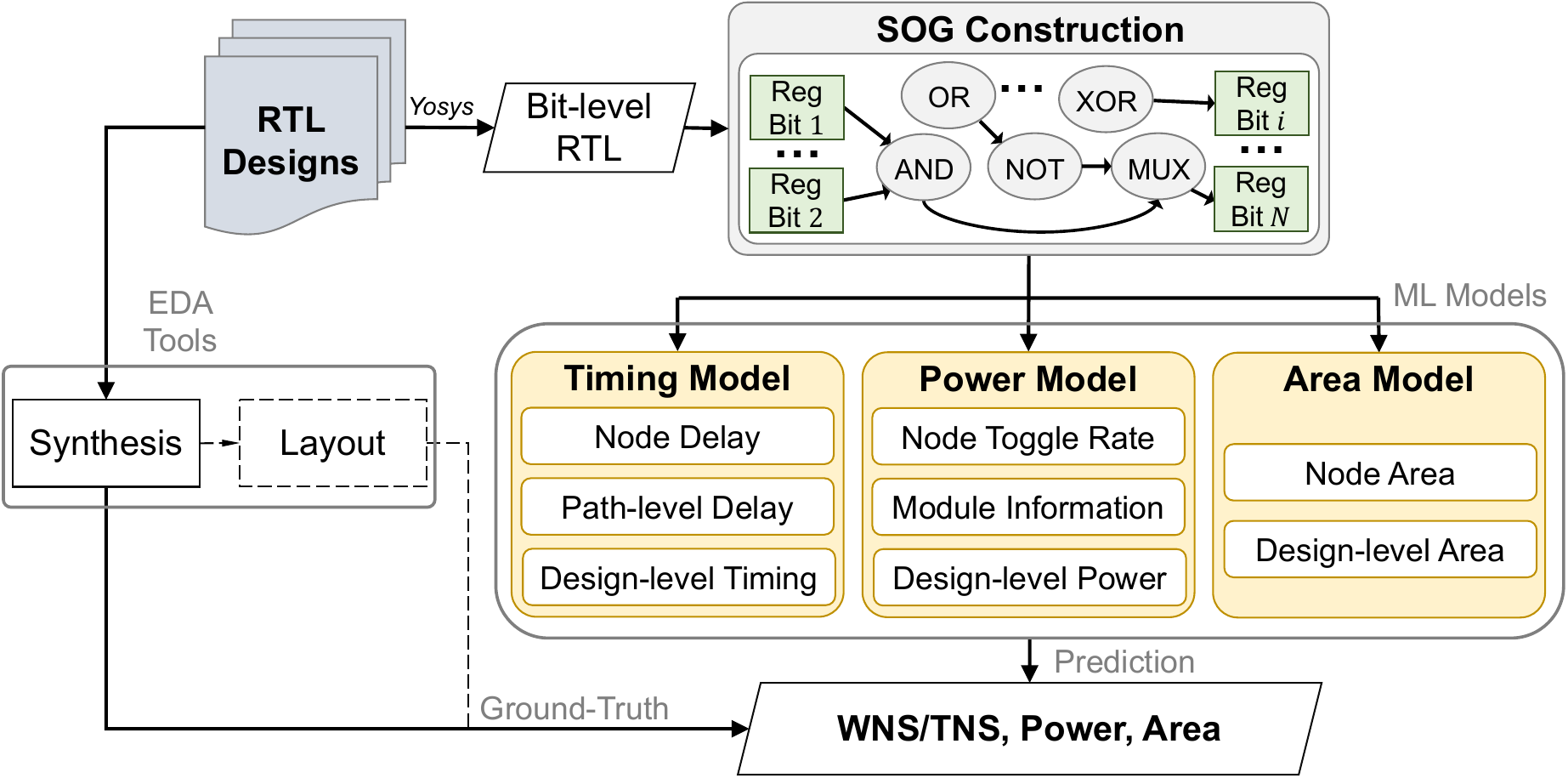}
  \caption{MasterRTL overall workflow for RTL-stage design PPA prediction. MasterRTL utilizes a general SOG representation to capture the design RTL, while exploiting customized models for timing, power, and area to achieve precise design quality prediction.}
  \label{fig:workflow}
  \vspace{-.1in}
\end{figure}

\section{Methodology} \label{sec:algorithm}

% as summarized in Fig.~\ref{fig:workflow} and Equation~\ref{eq:formulation}

% for estimating $\{P_\mathcal{G}, T_\mathcal{G}, A_\mathcal{G}\}$, respectively

This section introduces our MasterRTL framework in detail. 
Let's denote an RTL design in  HDL-code format as $\mathcal{H}$, and the generated gate-level netlist after synthesis as $\mathcal{G}$, with its  power, timing, and area as $\{P_\mathcal{G}, T_\mathcal{G}, A_\mathcal{G}\}$.
%Denote the original HDL-code format of an RTL design as $\mathcal{H}$. After synthesis, denote the generated gate-level netlist as $\mathcal{G}$, the power, timing, and area of the netlist $\mathcal{G}$ as $\{P_\mathcal{G}, T_\mathcal{G}, A_\mathcal{G}\}$. 
The target of our RTL modeling framework $F$ is to evaluate these post-synthesis qualities of any RTL design. This framework first converts HDL-code format $\mathcal{H}$ to a representation $\mathcal{R}$, allowing the processing of RTL details. Then different power, timing, and area models $\{f_p, f_t, f_a\}$ will be developed separately. The target can be summarized as below.
\begin{equation}
    F(\mathcal{H}) = \{f_p(\mathcal{R}), f_t(\mathcal{R}), f_a(\mathcal{R})\} \ \rightarrow \  \{P_\mathcal{G}, T_\mathcal{G}, A_\mathcal{G}\} \label{eq:formulation}
\end{equation}

% of this representation $\mathcal{R}$
In this section, we will first introduce the new RTL representation $\mathcal{R}$ named SOG adopted by MasterRTL, together with a preview of experimental results demonstrating its advantages. Based on this new representation $\mathcal{R}$, we will introduce the new timing, power, and area models proposed in MasterRTL. % separately. 

\subsection{SOG: Our Suggested Bit-Level RTL Representation}
\label{sec:representation}
\begin{figure}[!t]
  \centering
  \includegraphics[width=1\linewidth]{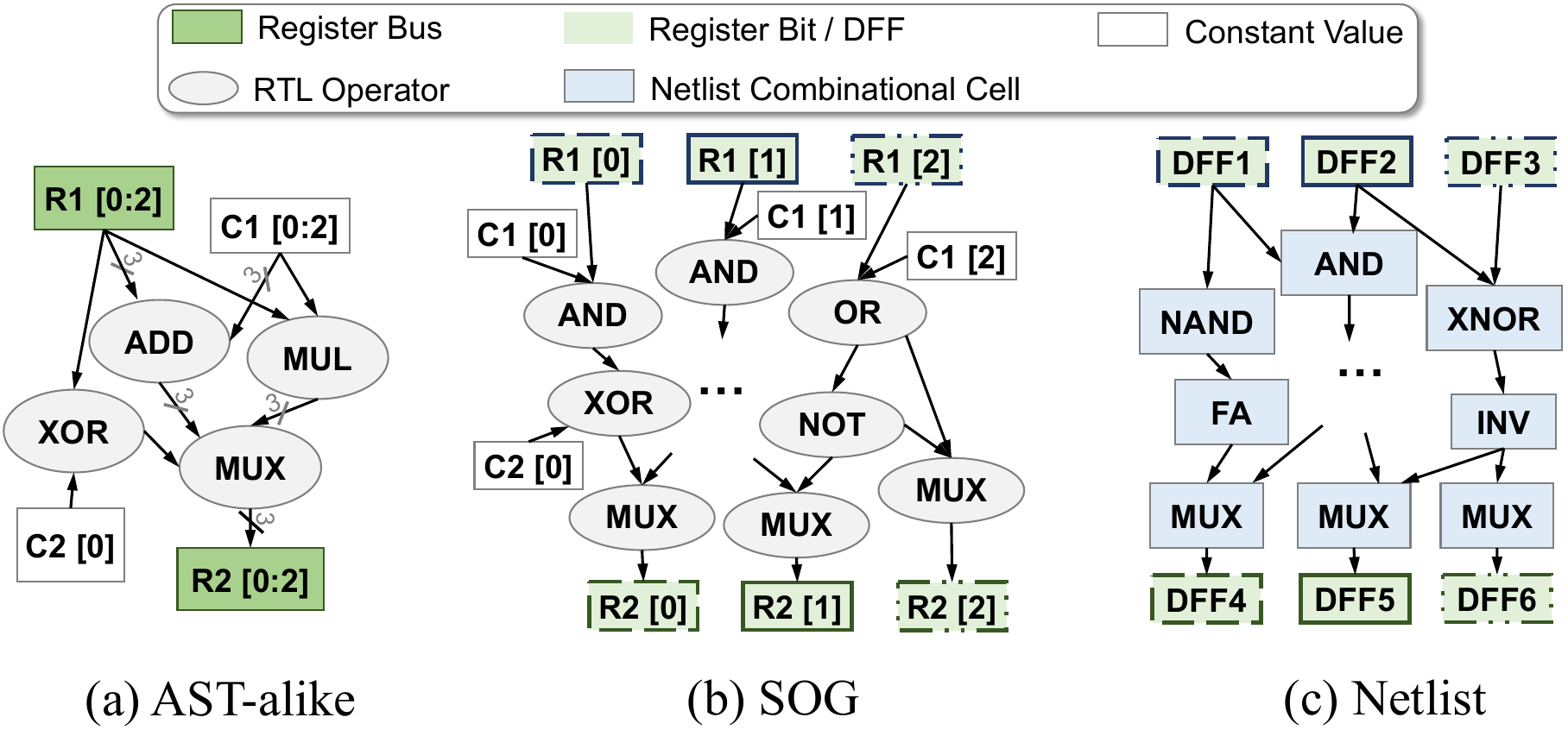}
  \caption{Comparison between different RTL representations and the target gate-level netlist. (a) Abstract syntax tree (AST)-alike RTL representations with each node denoting word-level design operations, adopted in prior works~\cite{xu2022sns, sengupta2022good}. (b) Our adopted bit-level RTL representation named simple operator graph (SOG). With only 5 operator types, it proves to be significantly more general across different designs and more similar to the target gate-level netlist. (c) The ultimate target gate-level netlist after logic synthesis. 
  Registers exhibit 1-to-1 matching between SOG and netlist.
  }
  \label{fig:reper}
  \vspace{-.1in}
\end{figure}

The RTL-stage PPA modeling starts with converting the raw HDL code $\mathcal{H}$ to a reasonable data structure, named design representation $\mathcal{R}$, %so that we can process RTL design information.
to enable the processing of detailed RTL design information. A major challenge in this task is how to bridge the huge gap between the RTL-stage representation and the post-synthesis gate-level netlist, without invoking the time-consuming logic synthesis engine. Prior works~\cite{xu2022sns, sengupta2022good} directly convert the HDL code to an AST-alike representation. As Fig.~\ref{fig:reper}(a) shows, this AST-alike representation can be viewed as a directed graph, with each node being a word-level design operation with any bit width. These operations may include adders, subtractors, multipliers, shifters, comparators, multiplexers, registers, and other logic operators~\cite{xu2022sns}. Altogether there are 18 different word-level operations with any legitimate bit width in such AST-alike representation. Fig.~\ref{fig:reper}(c) shows the target gate-level netlist after synthesizing this RTL example.

%. We thus refer them as AST-alike word-level representation. 
% fast conversion from the word-level representation to a
% \yao{How the bit-level representation is generated.} 

%In this work, MasterRTL adopt a new bit-level RTL representation, without any time-consuming optimization in the existing synthesis process. We name it simple-operator graph (SOG), which is shown in Fig.~\ref{fig:reper}\subref{fig:reper_b}. Compared with the prior AST-alike representation in Fig.~\ref{fig:reper}\subref{fig:reper_a}, SOG breaks a multi-bit word into logic bits and replaces a word-level operator with the corresponding Boolean relations over these bits.
%The replacement follows a pre-set mapping relation for each word-level operator. This bit-level representation can be generated based on open-source tools like Yosys~\cite{wolf2013yosys} and finishes within seconds. Compared with 18 different word-level operations in AST-alike representation, the bit-level representation SOG consists of only 6 types of single-bit primary simple logic operations, including two-input AND, two-input OR, two-input XOR, NOT, 2-to-1 MUX, and single-bit register. 

In comparison, MasterRTL adopts a new bit-level RTL representation without involving time-consuming optimizations in the existing synthesis process. It is shown in Fig.~\ref{fig:reper}(b). Compared with 18 different word-level operations in AST-alike representation, this new bit-level representation $\mathcal{R}$ consists of only single-bit registers and 5 types of single-bit primary simple logic operations, including two-input AND, two-input OR, two-input XOR, NOT, 2-to-1 MUX. Therefore, we name it simple operator graph (SOG). This new bit-level RTL representation SOG is generated by breaking each multi-bit word into logic bits and replacing the word-level operations with corresponding Boolean relations over these bits, following a pre-set mapping relation. The generation of this SOG representation can be implemented based on open-source tools like Yosys~\cite{wolf2013yosys} and finishes in a short time.

%In this subsection, we discuss the advantage of SOG in ML-based prediction, compared with the AST-like representation in prior works. We will start with intuitive observations on these two formats, then summarize our observed advantages, followed by a preview of our experiment results to support our claims. 

%By observing Fig.~\ref{fig:reper}, the bit-level representation in Fig.~\ref{fig:reper}\subref{fig:reper_b} is intuitively more similar to the target netlist in Fig.~\ref{fig:reper}\subref{fig:reper_c}, in terms of graph scale and granularity. In addition, 

In this subsection, we inspect the advantage of SOG for ML predictions, compared with AST-alike representation in prior works~\cite{xu2022sns, sengupta2022good}. %By only adopting 5 simple single-bit logic operations, the SOG shown in Fig.~\ref{fig:reper}(b) can better unify different design types and styles. 
We observe two potential advantages of SOG as follows. 
%  Actually, these intuitively observations have been validated in our experiments. 
%After our detailed analysis, we propose two key advantages of the bit-level representation, as summarized below. 
%The first advantage is, the bit-level representation correlates better with the target gate-level netlist after logic synthesis. 
%Second, the bit-level representation is more general, leading to less difference among defferent design types and styles. As a result, this bit-level representation is more ML-friendly for early-stage evaluation tasks. We will provide two relevant observations below. 
\begin{enumerate}
\item \emph{Similarity}: SOG is more similar to the target gate-level netlist after logic synthesis. It reduces the gap between pre-synthesis RTL and post-synthesis gate-level netlist. 
\item \emph{Generalization}: By only adopting 5 simple single-bit logic operations, SOG is more general than the AST, reducing the difference among various RTL design types and styles.
\end{enumerate}

%with more detailed analysis will be given in Section. 
% evaluation on these two advantages

%The first \emph{similarity} advantage of bit-level representation can be first intuitively reflected by the toy example in Fig.~\ref{fig:reper}, where the bit-representation in Fig.~\ref{fig:reper}\subref{fig:reper_b} is more similar to the netlist in Fig.~\ref{fig:reper}\subref{fig:reper_c}. 

To support these two claims, we provide a preview of our experiment results in Table~\ref{tbl:summary_observe}, which compares the area prediction accuracy based on two different representations $\mathcal{R}$. It is based on our dataset with 90 different designs. Here we present the area prediction task considering its simplicity.
The trend is similar in power and timing predictions shown by full experimental results in Section~\ref{sec:results}.
%as we will introduce in our experimental results in Section~\ref{sec:results}, the trend is similar in power and timing predictions. 

\begin{table}[!t]
      \centering
      \renewcommand{\arraystretch}{1.1}
      \resizebox{0.49\textwidth}{!}{
        \begin{tabular}{ |c|c||c|c|c|c|c||c| } 
        \hline
    %      \multicolumn{2}{|c|}{\multirow{2}{*}{\textbf{Area Prediction Accuracy}}}   & \multicolumn{2}{c|}{Area Prediction Accuracy}\\
     %   \cline{3-4}
         \multicolumn{2}{|c||}{\multirow{2}{*}{Area Estimation Accuracy}}   &  \textbf{SOG}  & AST-alike \\
         \multicolumn{2}{|c||}{}    &  (bit-level)  & (word-level) \\
        \hline
        \hline
         \multicolumn{2}{|c||}{Simple correlation of all designs} &  $R=0.976$ & $R=0.862$  \\
         \hline
         \multicolumn{3}{c}{}\\[-.12in]
         \multicolumn{4}{c}{(a) Evaluation of similarity between RTL representations and netlist.}  \\[-.05in]
         \multicolumn{3}{c}{}\\
         \hline
            \multicolumn{2}{|c||}{Area Estimation Accuracy}  & \textbf{SOG}  & AST-alike \\
        \cline{1-2}
        %\hline
           Training data   &  Test data &  (bit-level)  & (word-level) \\
        \hline
         \hline
        Random & Other Random  & $R=0.98$   &$R=0.94$ \\
         \hline
         %CPUs & Non-CPU Designs &  $R=0.969$  & $R=0.208$  \\
         CPUs & Non-CPU designs &  $R=0.97$  & $R=0.21$  \\
         \hline
         %Large design & small design & $R=0.942$ & $R=0.153$   \\
         Large designs & Small designs & $R=0.94$ & $R=0.15$   \\
        \hline
         \multicolumn{3}{c}{}\\[-.12in]
         \multicolumn{4}{c}{(b) Evaluation of the generalization ability of RTL representations.}  \\
        \end{tabular}  % Many
       }
        \caption{Comparison between SOG (with bit-level operators) and AST-alike representations (with word-level operators)  using the straightforward area prediction task. (a) The average area correlation between RTL representation and gate-level netlist. SOG gives obviously better correlation, indicating a significantly better \emph{similarity} with netlist. (b) Area prediction test accuracy when using different training/testing designs. Only SOG remains accurate for all three scenarios, implying a significantly more \emph{general} representation.}%while AST-alike representation' accuracy significantly degrade when different design types are used for training and testing. } 
        \label{tbl:summary_observe}
        \vspace{-.05in}
\end{table}

Table~\ref{tbl:summary_observe}(a) shows a higher \emph{similarity} between the SOG representation and the netlist. It simply measures the Pearson correlation $R$ between extracted features from each RTL representation and the gate area of netlist as ground-truth, without any explicit ML model. Extracted features for each representation are the numbers of node operations in each design. There are 6 features for SOG and 18 features for AST-alike representation. Despite a smaller number of features, SOG correlates much better with netlists, according to Table~\ref{tbl:summary_observe}(a). Such a better correlation implies there is a higher similarity between SOG and the netlist.

%  on two different representations

Table~\ref{tbl:summary_observe}(b) implies that SOG is a more \emph{general} representation. It reports the test accuracies of our proposed simple area model, which will be introduced in subsequent sections. When we randomly assign the 90 designs to the training and testing dataset, as Table~\ref{tbl:summary_observe}(b) shows, the model based on SOG is slightly better than the AST-alike representation, again reflecting SOG's better similarity with netlist. However, when we use distinct types of designs for training and testing (e.g., train on CPU/large designs while test on non-CPU/small designs), the accuracy of models using AST-alike representation immediately degrades from $R=0.94$ to $R=0.21/0.15$. In comparison, ML models based on SOG remain accurate with $R=0.97/0.94$. These results imply that SOG is a more \emph{general} representation that unifies different designs in training and testing datasets.\looseness=-1

Besides the above two advantages in \emph{similarity} and \emph{generalization}, there is a third key advantage of SOG named \emph{consistency} that enables the development of our new customized models in MasterRTL: 
\begin{enumerate}
\setcounter{enumi}{2}
    \item \emph{Consistency in register mapping}: As Fig.~\ref{fig:reper} shows, every register cell in gate-level netlist $\mathcal{G}$ has a 1-to-1 mapping to the single-bit register operator in SOG $\mathcal{R}$. In comparison, the AST with word-level nodes can never directly map to netlists in prior works. %every single-bit register operator in SOG can be mapped directed to its corresponding real register cell in gate-level netlist. 
\end{enumerate}
%This 1-to-1 mapping between SOG representation and netlist enabled the development of our new timing, power, and area models. 
%Supported by this 1-to-1 register mapping, 
Supported by the consistency in 1-to-1 register mapping, our significantly more accurate and detailed timing model can directly explore critical paths on SOG, then map to paths of the same starting and end registers on netlist. %For our new power model, we can annotate toggle rate on each register node of SOG. 
In the next subsections, we will separately introduce our timing model, power model, and area model in detail.

\subsection{RTL-Stage Timing Modeling}

\begin{figure*}
  \centering
  \includegraphics[width=0.9\linewidth]{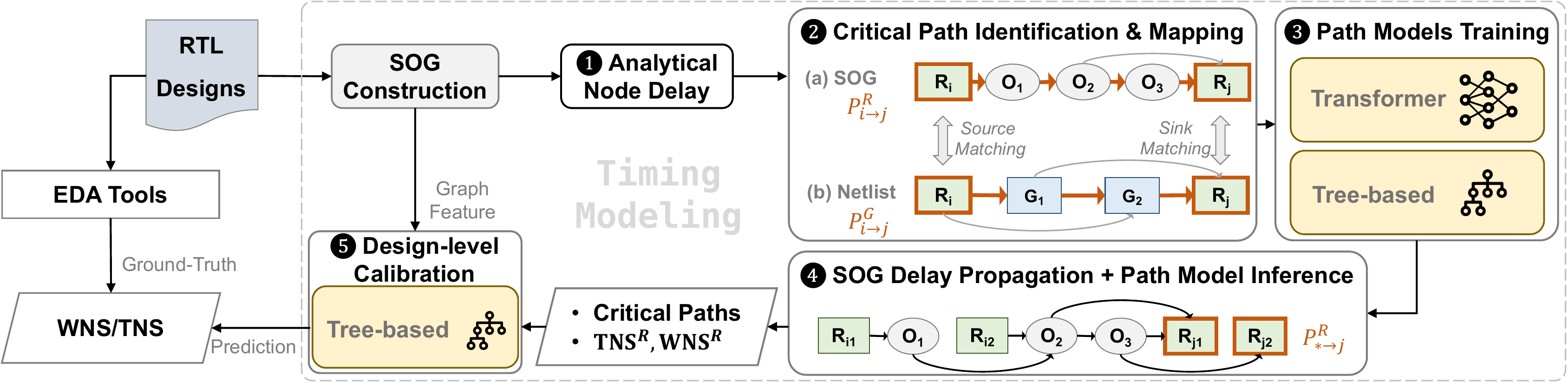}
  \caption{Timing evaluation flow in MasterRTL. The timing models capture all critical paths at RTL stage, enabling accurate timing estimation.}
  \label{fig:timingflow}
\end{figure*}

For the RTL-stage timing modeling, as Fig.~\ref{fig:timingflow} shows, we propose a multi-stage ML framework to evaluate both the TNS and WNS of any RTL design. The target ground-truth timing information including TNS and WNS values are from the post-synthesis timing report. Please notice that detailed timing evaluation at such an early stage is highly challenging, since neither logic optimization nor technology mapping has been performed yet. Therefore, different from previous netlist- or layout-stage timing modeling methods~\cite{xie2021timing, barboza2019machine, guo2022timing}, our modeling method will primarily focus on key patterns and use some approximations. While it may result in imperfections, they will be addressed by subsequent calibration using ML models. \looseness=-1

For RTL-stage timing modeling, a key challenge is that the ground-truth label is based on the timing analysis on gate-level netlist $\mathcal{G}$. Such netlist-level timing values cannot be directly mapped to the AST-alike RTL representation $\mathcal{R}$. Therefore, state-of-the-art works either give up modeling RTL details~\cite{sengupta2022good} or only use synthesized pseudo paths as training data~\cite{xu2022sns}, leading to significant inaccuracies. In comparison, MasterRTL utilizes the consistency of registers between the new SOG representation $\mathcal{R}$ and netlist $\mathcal{G}$. 
Specifically, we will capture the slowest critical paths in both SOG representation $\mathcal{R}$ and netlist $\mathcal{G}$, and map these paths one-by-one according to their starting and end registers cell/operator. %We will train path-level models based on these paths. %we will first capture the slowest critical paths between any pairs of registers, then train path-level ML models with ground-truth critical path delay between the identified pair of registers. 
%Then TNS and WNS values can be predicted by aggregating path delays with a last calibration ML model. 
This multi-stage timing modeling process is introduced in detail below. 

%by adopting the SOG representation $\mathcal{R}$, 
%before and after the synthesis process. 

%unavoidable errors will be . 
%more efficient light-weight models. 
% The delay value simply depends on the driving operation and the number of fan-outs. More precisely, the simple
% We assume a fixed input slew value in this initial step, in order to avoid the time-consuming slew propagation across operations. 
%  of the target technology node

% that is proportional to
\circled{1} \textbf{Node-delay modeling in $\mathcal{R}$.}  We first develop a simple approximated analytical (non ML-based) model to evaluate the delay after each node on our RTL representation $\mathcal{R}$ in SOG. Please notice that such `node-delay' on $\mathcal{R}$ is not intended to be a real delay value. Instead, we only use it to guide path extraction on $\mathcal{R}$ for feature collection. This node-delay model is a linear function of the fan-out number, with coefficients determined by the type of the driving node operator. Specifically, for each type of node operator, its coefficients are approximated by the RC values of standard cells of the same type in the liberty file (e.g., .lib/.db).

\circled{2} \textbf{Critical path identification and mapping in $\mathcal{R}$ and $\mathcal{G}$.}
%Critical path means the path with the maximum path delay or arrival time. 
Based on the estimated node-level delay in $\mathcal{R}$, we will efficiently propagate such estimated node-level delay by traversing the SOG graph in topological order, in order to estimate the %most critical 
maximum-delay path $P^{\mathcal{R}}_{i \rightarrow j}$ between any pair of registers $i \rightarrow j$ in the representation $\mathcal{R}$, with the $i^{\text{th}}$ register as the startpoint and $j^{\text{th}}$ register as the endpoint. For a design with $N$ registers,  $i, j \in [1, N]$. 

Based on the captured $P^{\mathcal{R}}_{i \rightarrow j}$ from RTL representation $\mathcal{R}$, the goal is to evaluate ground-truth maximum path delay between the same pair of registers in the gate-level netlist $\mathcal{G}$. Denote the target critical path between register $i$ and $j$ in the netlist as $P^{\mathcal{G}}_{i \rightarrow j}$. Its path delay is collected using post-synthesis timing analysis EDA tools.

%Based on the captured critical paths $P^{\mathcal{R}}_{i \rightarrow j}$ from RTL representation $\mathcal{R}$, the next goal is to evaluate ground-truth maximum path delay between the same pair of registers in the gate-level netlist $\mathcal{G}$. Denote the target critical path between register $i$ and $j$ in the netlist as $P^{\mathcal{G}}_{i \rightarrow j}$. Its path delay is collected using post-synthesis timing analysis EDA tools. 

% For a design with $N$ registers,  $i, j \in [1, N]$. Critical path means the path with the maximum path delay or arrival time.
% Based on the captured critical paths $P^{\mathcal{R}}_{i \rightarrow j}$ from RTL representation $\mathcal{R}$, the timing model will evaluate of the ground-truth maximum path delay between the same pair of registers in the gate-level netlist $\mathcal{G}$.  

%captured with node delay i the previous step, while a node in $P^{\mathcal{G}}_{i \rightarrow j}$ represents a standard cell, captured by timing analysis tools.  
%A node in $P^{\mathcal{R}}_{i \rightarrow j}$ represents an operator captured with node delay i the previous step, while a node in $P^{\mathcal{G}}_{i \rightarrow j}$ represents a standard cell, captured by timing analysis tools.  

Notice that both critical paths $P^{\mathcal{R}}_{i \rightarrow j}$ and $P^{\mathcal{G}}_{i \rightarrow j}$ share the same startpoint and endpoint registers $\{i, j\}$, but the actual nodes on each path are different. A mapping example is shown in Fig.~\ref{fig:timingflow}, where a node in $P^{\mathcal{R}}_{i \rightarrow j}$ represents an RTL operator, while a node in $P^{\mathcal{G}}_{i \rightarrow j}$ represents a standard cell. 

\circled{3} \textbf{Path-level delay model training.}  Based on features from the RTL-stage path $P^{\mathcal{R}}_{i \rightarrow j}$, we train a path-level model $f_t^{\text{path}}$ to estimate the ground-truth path delay label at netlist $P^{\mathcal{G}}_{i \rightarrow j}$. 
\begin{equation}
    f_t^{\text{path}} (P^{\mathcal{R}}_{i \rightarrow j}) \ \rightarrow \ \text{The path delay of } P^{\mathcal{G}}_{i \rightarrow j}
\end{equation}
To build the training dataset, for each design, we identify the register pairs $\{i, j\}$ that lead to the top $1\%$ maximum-delay netlist paths $P^{\mathcal{G}}_{i \rightarrow j}$, according to the post-synthesis timing report. %Then path delay of $P^{\mathcal{G}}_{i \rightarrow j}$ is recorded. 

For this path-level model $f_t^{\text{path}}$, we explored two types of ML models. The first is the popular transformer model adopted in large language models (LLMs) nowadays, processing each path as a sequence of nodes. This transformer model is similar to the one adopted in~\cite{xu2022sns}. But due to its limitation of using individual pseudo training paths, \cite{xu2022sns} has no fan-out/degree information in features, which are critical for path delay prediction. In comparison, we incorporate the fan-out count of each node (i.e., number of child nodes) in the path $P^{\mathcal{R}}_{i \rightarrow j}$ into the input features. 
The second explored ML model type for $f_t^{\text{path}}$ is the traditional tree-based algorithms like Random Forest~\cite{breiman2001random} with careful feature engineering. The features captured from the path $P^{\mathcal{R}}_{i \rightarrow j}$ include: 1) the total number of all operations; 2) the number of each type of operation; 3) the accumulated node delay on this path according to the model in \circled{1}. % Then the estimated slack of each critical path between registers $i \rightarrow j$ equals the clock period subtracted by the estimated path delay.

% Then leveraging the analytical node delay from the first step, 
% Then we propagate the estimated node-level delay mentioned in the first step, by traversing the SOG graph at the topological order.

\circled{4} \textbf{Path-level delay model inference.}  After training the path-delay model $f_t^{\text{path}}$, it will be applied for TNS/WNS prediction on any given new RTL design. Similar to the calculation process of TNS and WNS in netlists, in the representation $\mathcal{R}$, we capture the critical path of each register as the endpoint. Using the technique in \circled{2}, we propagate the estimated node delays by traversing the SOG. Consequently, we identify altogether $N$ paths, with each path denoted as $P^{\mathcal{R}}_{* \rightarrow j}$ ($j \in [1, N]$). Here $*$ represents the startpoint register that leads to the maximum path delay.

Then the trained path-level model predicts the path delay of all $N$ paths $P^{\mathcal{R}}_{* \rightarrow j}$. %, mitigating the path delay gap between SOG and netlist. 
According to the definition of TNS and WNS, their estimations in the representation $\mathcal{R}$ are calculated below. 
\begin{align*}
    \text{TNS}^{\mathcal{R}} &= \sum_{j=1}^{N} (clk - f_t^{\text{path}}(P^{\mathcal{R}}_{* \rightarrow j}))  \\
    \text{WNS}^{\mathcal{R}} &= \min_{j \in [1, N]} (clk - f_t^{\text{path}}(P^{\mathcal{R}}_{* \rightarrow j}))
\end{align*}

%For each path, we extract features based on each node in the path and its fan-out nodes. The label here is the actual critical path's arrival time in the post-synthesis design. Then the estimated slack of each path is the estimated arrival time minus the clock period. 

\iffalse
\begin{figure}[htbp]
	\centering
	\subfloat[Small]{\includegraphics[height=0.51\linewidth, max width=0.38\linewidth]{_fig/hist_s.pdf}}\hspace{0pt}
	\subfloat[Medium]{\includegraphics[height=0.51\linewidth, max width=0.38\linewidth]{_fig/hist_m.pdf}}\hspace{0pt}
	\subfloat[Large]{\includegraphics[height=0.51\linewidth, max width=0.38\linewidth]{_fig/hist_l.pdf}}\hspace{0pt}
	\caption{Hist}
        \label{fig:hist}
\end{figure}
\fi

\begin{figure}[!b]
	\centering
 \vspace{-.1in}
 \hspace{-.05in}
	\subfloat[Small Designs]{\includegraphics[height=0.58\linewidth]{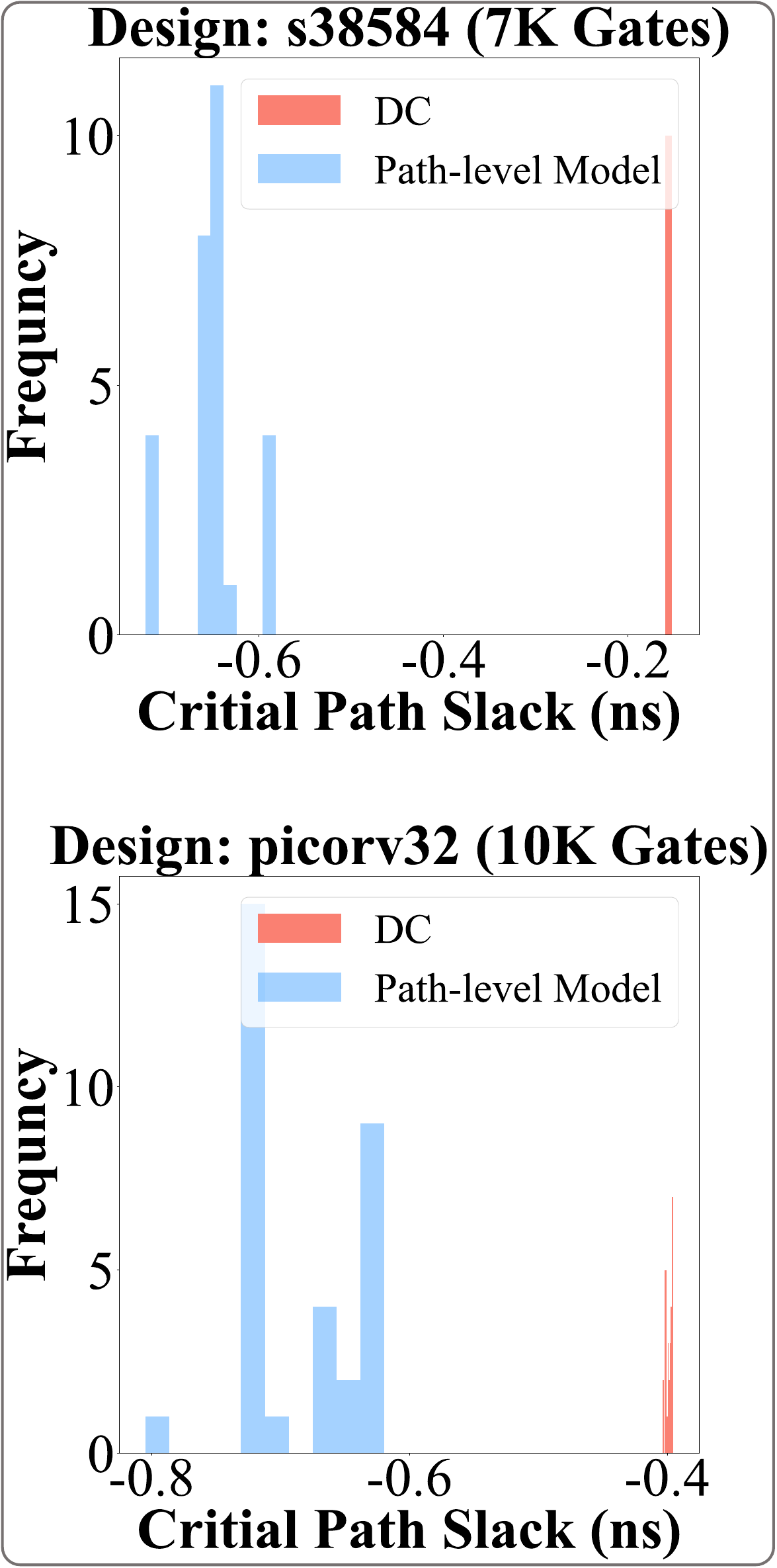}}\hspace{.05in}
	\subfloat[Medium Designs]{\includegraphics[height=0.58\linewidth]{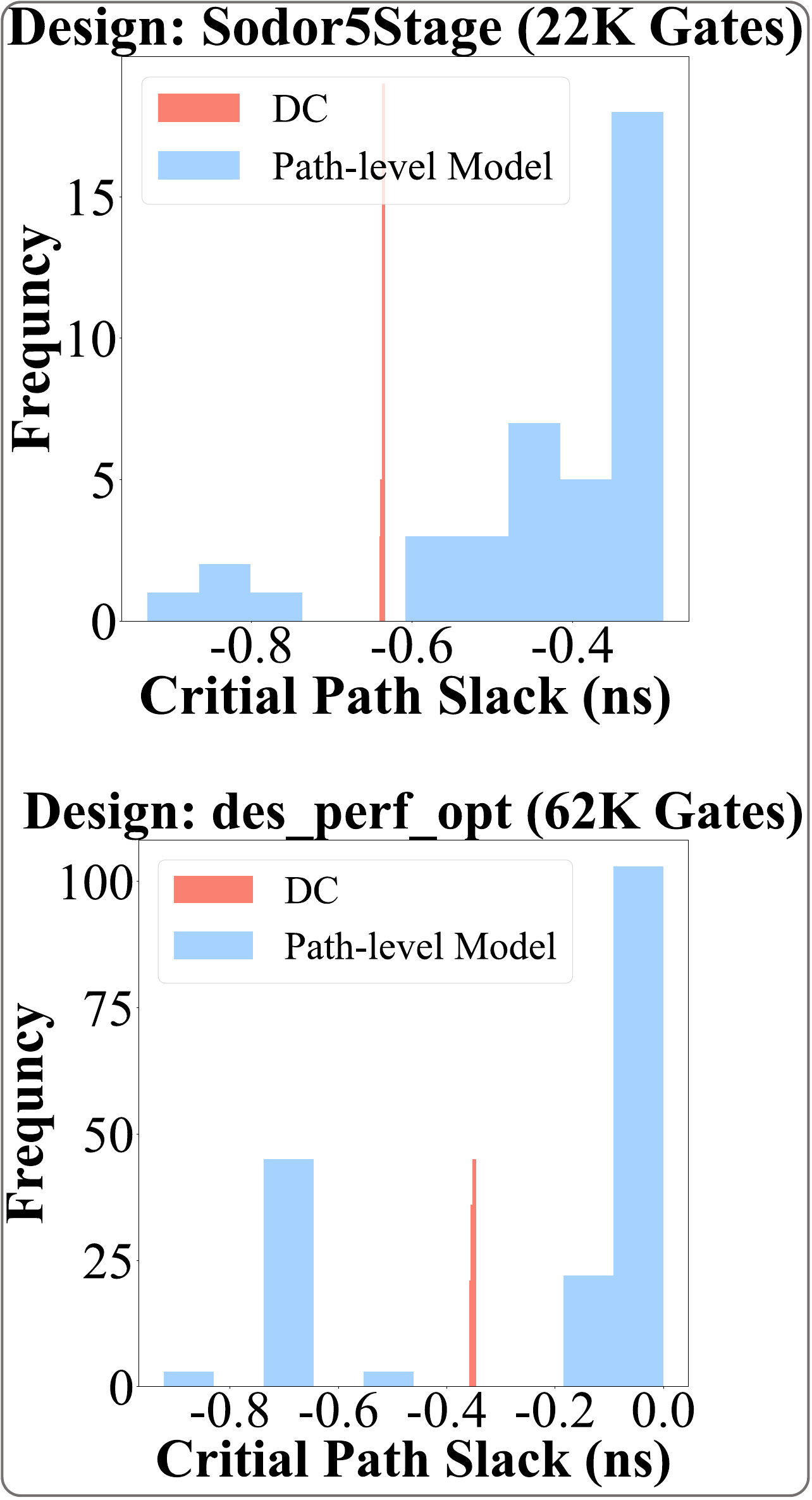}}\hspace{.05in}
	\subfloat[Large Designs]{\includegraphics[height=0.58\linewidth]{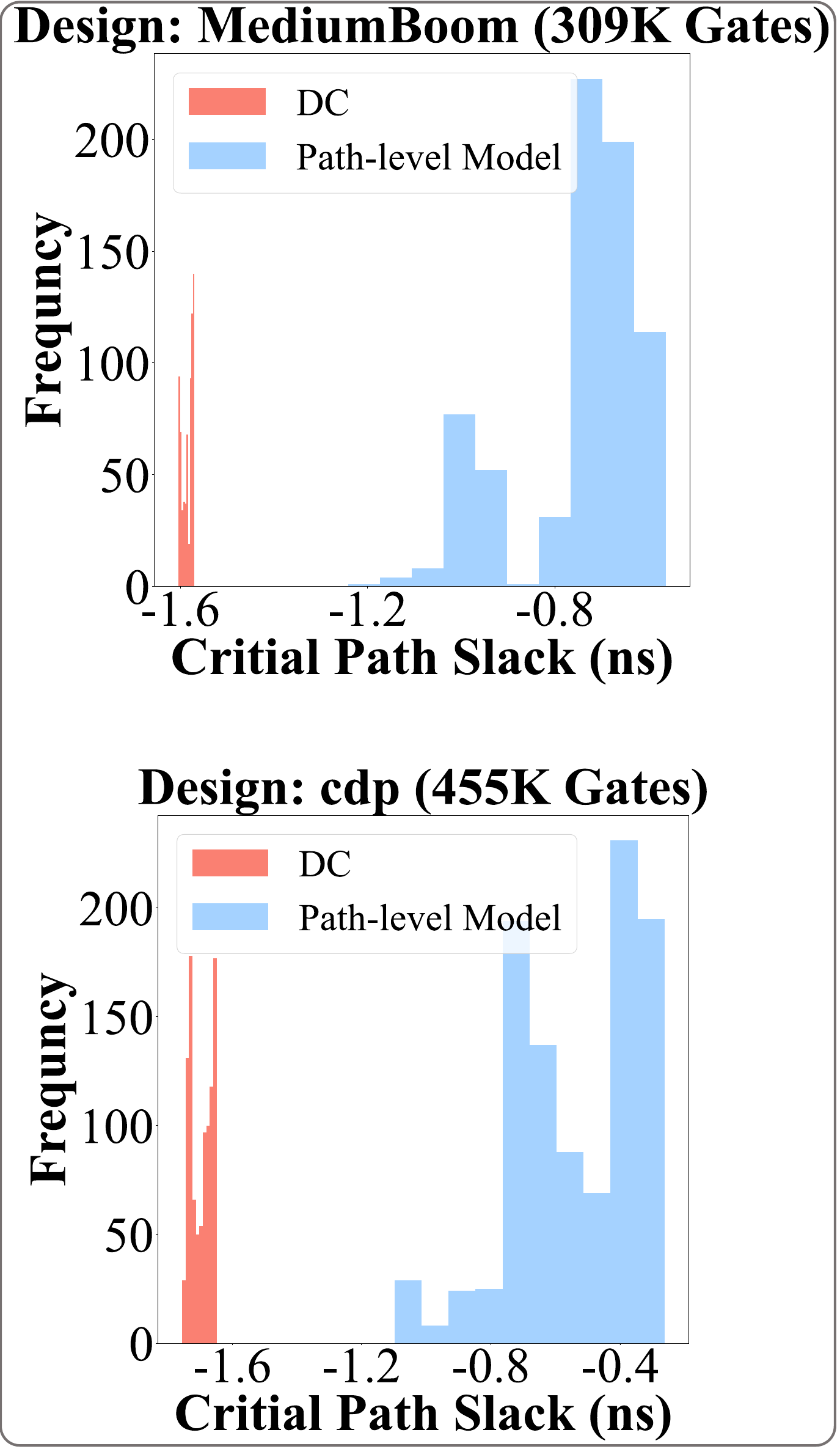}}\hspace{0pt}
	\caption{The path delay distribution of the worst 1\% critical paths from both netlist (DC) and our path-level model. Although their distributions are not exactly the same, there are obvious patterns depending on the design size. }
        \label{fig:hist}
\end{figure}

\circled{5} \textbf{Design-level TNS/WNS calibration.} 
Given that  RTL-stage timing modeling is highly challenging,
%Considering the high challenge of RTL-stage timing modeling, 
the estimated path delays, WNS$^{\mathcal{R}}$, and TNS$^{\mathcal{R}}$ predicted directly by the path-level model in \circled{4} are not sufficiently accurate. But they provide useful information for further calibration. %we observed a significant disparity in critical path slack between the calibrated paths and the target netlist. 
In Fig.~\ref{fig:hist}, we compare the slack distribution of the worst 1\% of the $N$ critical paths from both our path-level model prediction $clk - f_t^{\text{path}}(P^{\mathcal{R}}_{* \rightarrow j})$ and ground-truth gate-level netlist timing report. We observe several % highly 
interesting and reasonable patterns in Fig.~\ref{fig:hist}:\looseness=-1

\begin{itemize}
    \item The distribution of critical paths (red) from netlists is more concentrated than the path-level model predictions (blue).
    \item While there are discrepancies between the ground-truth and path-level predictions, clear patterns exist based on design size. Specifically, predictions on small-scale designs tend to be over-pessimistic, large-scale ones are over-optimistic, and medium-sized designs fall in the middle.
\end{itemize}

%According to our observation, %iterative and distinct %by EDA tools
These discrepancy patterns are actually expected. They primarily attributed to the optimization efforts carried out during logic synthesis, specifically targeting the most critical paths. Such optimization efforts make slacks of the 1\% critical paths in netlist $\mathcal{G}$ more concentrated and closer to actual WNS. Also, the impact of these optimization efforts is more obvious in small designs with fewer paths, making the path-level predictions based on $\mathcal{R}$ over-pessimistic by comparison, and vice versa.

%slack values better than predictions. 
%can better optimize .
%Therefore, after the synthesis, the slacks of the 1\% worst netlist critical paths are. 

Leveraging these valuable patterns, we further devote one additional final-stage model to calibrate the above evaluations TNS$^{\mathcal{R}}$ and WNS$^{\mathcal{R}}$ towards the actual TNS and WNS labels measured in gate-level netlist $\mathcal{G}$. It adopts a tree-based ML model, with features including: 1) the SOG features (i.e., the numbers of node operators) indicating the design scale; 2) the estimated TNS$^{\mathcal{R}}$ and WNS$^{\mathcal{R}}$ provided by the path-level model; 3) the slack distribution of the worst 1\% of the $N$ critical paths based on path-level model prediction $clk - f_t^{\text{path}}(P^{\mathcal{R}}_{* \rightarrow j})$. For each design, we take the worst, 10\%, 50\%, and 90\% percentile of the predicted slacks as features.

% 2) distribution of predicted path slacks on the worst $1\%*N$ paths (i.e., the worst, 10\%, 50\%, and 90\% percentile predicted slacks );

% obtained from the worst 1\% critical paths
% four values indicating the

%This final stage model aims to calibrate the above evaluations TNS$^{\mathcal{R}}$ and WNS$^{\mathcal{R}}$ towards the actual TNS and WNS labels measured in gate-level netlist $\mathcal{G}$. 

% In this last step, we will evaluate the ground-truth TNS and WNS of the target design. 

% Similar to the calculation process of TNS and WNS, in the representation $\mathcal{R}$, we capture the critical path of each register as the endpoint, resulting in altogether $N$ paths, with each path denoted as $P^{\mathcal{R}}_{* \rightarrow j}$ ($j \in [1, N]$). The $*$ means the startpoint can be any register that leads to the maximum path delay. According to the definition of TNS and WNS, their estimations should be calculated below. 
% \begin{align*}
%     \text{TNS}^{\mathcal{R}} &= \sum_{j=1}^{N_{Reg}} (clk - Delay(P^{\mathcal{R}}_{* \rightarrow j}))  \\
%     \text{WNS}^{\mathcal{R}} &= \min_{j \in [1, N_{Reg}]} (clk - Delay(P^{\mathcal{R}}_{* \rightarrow j}))
% \end{align*}

% In this step, we devote one additional tree-based ML model, with the features extracted from the RTL graph, to calibrate the above evaluations TNS$^{\mathcal{R}}$ and WNS$^{\mathcal{R}}$ towards the actual TNS and WNS labels measured in gate-level netlist $\mathcal{G}$. 

\subsection{RTL-Stage Power Modeling}

%Different from prior RTL-stage works~\cite{xu2022sns, sengupta2022good}, we are the first 

For RTL-stage power modeling, we propose a new toggle-rate-based and module-level method. %\hongce{We called this ``component-level'' in Sect. 1?}
Different from prior works~\cite{xu2022sns, sengupta2022good}, this is the first cross-design RTL-stage power model to utilize toggle rate as input. It annotates a toggle rate on every node of the SOG representation $\mathcal{R}$. %by traversing the graph. 
%Thanks to the simple node type in the adopted SOG, MasterRTL can easily annotate the toggle rate on all nodes in the SOG representation $\mathcal{R}$ by traversing the graph. 
%It first annotates the toggle rate on all $N_{Reg}$ registers, then propogates toggle rate to every nodes in our adopted SOG representation $\mathcal{R}$. 
Such toggle rate annotation supports both vector-based and vector-less power simulation scenarios.  
%depending on the analysis scenario.  %and post-synthesis power labels under specific workloads. per-cycle
For vector-based power analysis, we perform RTL simulations to obtain accurate toggle rates. It captures the dynamic behavior of the design, enabling precise estimation of toggle rates. % for each register.
For vector-less power analysis, we extract toggle rates from logic synthesis tools before the beginning of the synthesis process. It generates the necessary information about the design's toggling behavior without explicit workload simulation. 

%Leveraging these tools, we obtain the necessary information about the design's toggling behavior, facilitating toggle rate calculation without explicit simulation data.

% \begin{figure}
%   \centering
%   \includegraphics[width=1\linewidth]{_fig/power_flow.pdf}
%   \caption{Power evaluation flow in MasterRTL. The power models incorporate module-level RTL design partitioning and toggle rate information, enhancing power estimation accuracy. \yao{Perhaps add toggle rate generation here, or perhaps remove this fig}}
%   \label{fig:powerflow}
% \end{figure}
% Similar to the delay propagation algorithm in timing modeling

\textbf{Node toggle rate propagation.} 
The toggle rate information of all register bits in RTL can be obtained from the Switching Activity Interchange Format (SAIF) files from EDA tools for both vector-based and vector-less analysis. 
Then the toggle rate of all registers can be directly mapped to the register nodes on SOG. Starting from registers, we can efficiently propagate the toggle rate to every node on SOG $\mathcal{R}$ by traversing the graph, according to the functionality of each simple operator type in SOG~\cite{weste2015cmos}.

%(e.g., A CPU design can be decomposed into ALU, FPU, register files, etc.). 
% We propose a partitioning approach where the SOG graph is divided into N sub-graphs based on the hierarchical structure of Verilog code. 

\textbf{Module-level power estimation.}
A major challenge in design power prediction is the lack of training data, since each design only provides one data sample (i.e., its total power) for training. 
Therefore, instead of directly predicting the total power of an entire design, we break the design down into $M$ modules based on the RTL hierarchy of HDL code\footnote{Specifically, we partition modules by considering only one level below the top module in this work.}. 
This module-level method greatly enriches the total amount of power data by $M$ times. We predict the power of each module-level partition, denoted as $Power^{\mathcal{G}1}$, $Power^{\mathcal{G}2}$, ..., $Power^{\mathcal{G}M}$.
The overall power of the design, $Power^{\mathcal{G}}$, is then obtained by summing the power of each module-level partition: $Power^{\mathcal{G}} = \sum_{i=1}^{M} k_i \cdot Power^{\mathcal{G}i}$, where $k_i \in \mathbb{Z}^+$ refers to the number of times the $i^{th}$ module is instantiated.

%Specifically, we partition modules by considering only one level below the top module, to create separate Verilog files for each module. %and the remaining parts using Yosys. The ground-truth module-level labels are collected using EDA tools after the synthesis stage. This approach allows us to effectively analyze and model the power consumption at the module-level granularity.

% Instead of directly predict power $Power^{\mathcal{G}}$ on the entire graph, we partition the graph into N sub-graphs referring to the hierarchy of the modules in Verilog code, to predict each module-level total power ($Power^{\mathcal{G}1}$, $Power^{\mathcal{G}2}$, ... $Power^{\mathcal{G}N}$), where $Power^{\mathcal{G}} = \sum_{i=1}^{N} k_i \cdot Power^{\mathcal{G}i}$, and $k_i$ refers to the number of times the $i_th$ module is instantiated. 

% To be more specific, only modules under the top module will be segmented, and therefore, multiple module and the rest part in the top module will be generated into new Verilog files separately through Yosys, and number of times each module is instantiated will also be recorded. The ground truth labels of the modules are collected in EDA tools after synthesis.

For this module-level power prediction, we explored two types of models: graph neural network (GNN) model and traditional tree-based models. The total power is estimated by the sum of dynamic and static power. In terms of dynamic power, which is closely related to toggle rate, the explored models are introduced below. 
For the GNN model, we exploited the sub-SOG converted from modules as the input of the model. 
The features of each node are: %extracted through a graph traversal, which are: 
1) number of fan-in and fan-out; 2) one-hot encoding of the 6 node type; 3) propagated toggle rate. 
% we predict the total power by combining static and dynamic power, extracting features for these two components separately. In terms of dynamic power,
As for the tree-based method, for each module, we perform feature engineering based on toggle rate information: 1) the sum of toggle rate, 2) average toggle rates, 3) the sum of fan-out number multiples toggle rate on each node, 4) the total number of nodes. For static power estimation, it is toggle-rate irrelevant and only accounts for a small proportion of the total power, thus only the SOG-related features are utilized.\looseness=-1  

% A key feature is the sum of the static power of each node in the graph, according to the operator type from the liberty file. Additionally, we collect graph features on the sub-graph using the previously mentioned graph traversal method. 

\textbf{Design-level power calibration.} %Due to the different optimization efforts on distinct scales of designs during logic synthesis, 
Similar to the timing model, to capture the impact of different optimizations on distinct scales of designs during logic synthesis, we also add a final-stage tree-based ML model to calibrate the power prediction based on the sum of power prediction on all modules. 
In addition to this estimated total power, we incorporate the SOG graph features indicating the design scale. %features extracted from the SOG representation $R$. 
This final-stage calibration model further improves the accuracy.\looseness=-1

%By combining these features, we aim to refine and improve the accuracy of the post-synthesis total power prediction.

% Except for this estimated total power from previous step, we also combine the seven features from the graph.

\subsection{RTL-Stage Area Modeling}

According to our observation, area modeling is much more straightforward than the aforementioned timing and power modeling. Based on the SOG representation, a one-stage tree-based model is sufficient to provide high accuracy. %used to predict the post-synthesis area value. Similar to the total power estimation, 
The total gate area is decomposed into the area of sequential and combinational cells. 
%The ground truth labels for these components are obtained in the EDA tools. By leveraging this decomposition and utilizing a tree-based model, we are able to effectively predict the post-synthesis area of the design.
%\textbf{Design-level area prediction.} 
% utilize a simple calculation by
To predict the sequential area, we simply multiply the total number of registers in SOG with the cell area of a D-flip-flop in the liberty file. It provides accurate results, eliminating the need for further ML models. For the combinational part, we calculate the area of all operators in SOG using the liberty file, which is then combined with the previously mentioned SOG features to create a comprehensive feature vector. Based on the features, the tree-based model is further utilized for combinational area estimation.

%that adequately cover the training set. 
% We intend to develop this further in our future work, but here we provide a brief conceptual description of how this can be realized using the OPM. 

\subsection{Data Augmentation by Generating New RTL}
\label{sec:dataaug}
In real-world application scenarios, high-quality RTL designs are extremely important intellectual property (IP) of IC design companies and are hardly available to model developers. Therefore, it is very likely to encounter a shortage of diverse variants of RTL designs for model training. 
To address this data availability challenge, we present a new RTL data generation method. We intend to develop this further with more customizations on RTL format in our future work, but here we provide a brief description of how this can already be realized using the existing graph generation models.

Our current solution generates brand-new RTL designs based on a graph generative method~\cite{liao2019efficient} followed by a fine-tuning algorithm. %that enforces pre-defined RTL constraints (e.g., the proportion of each type of operator, bit-width, etc.). 
Specifically, we first convert a small number of available training RTL designs to the RTL representation\footnote{In this RTL generation task, we adopt AST-alike representation to train the graph generation model. As we discussed, SOG is closer to the gate-level netlist, while AST-alike representation is more similar to RTL.} in graph format, then train the graph generation model~\cite{liao2019efficient} with it. Then the graph generation model generates brand-new graphs. After that, the node type and connectivity in generated graphs will be fined-tuned to enforce our pre-defined RTL constraints (e.g., legitimate fan-in number, the proportion of each operator type, bit-width, etc.). Finally, the fine-tuned graphs can be converted back to newly generated RTLs as new training data for data augmentation. 

% the AST-alike graphs are further converted into pseudo RTL designs in Verilog. 

Currently, this RTL-stage data generation method is helpful when the available real RTL data is very limited, with its effect demonstrated in the Subsection~\ref{sec:grtl}. Considering the scaling-up trend of ML models, the serious data availability problem will become a key bottleneck in RTL modeling. Therefore, we believe this RTL generation method is highly promising, since it can generate an almost unlimited number of new RTL designs. In future work, we will try to remove the reliance on the extra fine-tuning algorithm and further improve the similarity between generated RTL designs and real ones.

 %  of this data augmentation method will be 

%This method aims to augment our existing dataset and enhance its variability, and a detailed demonstration will be given in Subsection~\ref{sec:grtl} 

\section{Experimental Results}
\label{sec:results}

\subsection{Experimental Setup and MasterRTL Implementation}
In this work, we first construct a comprehensive dataset by collecting altogether 90 different open-source RTL designs from different benchmark sources. Such a comprehensive dataset enables a more thorough examination of our proposed methodology. 

Table~\ref{tbl:benchmark} summarizes the sources of all designs we adopted in the dataset. They are originally coded with all mainstream HDLs, including Verilog, VHDL, Chisel~\cite{bachrach2012chisel}, and SpinalHDL~\cite{SpinalHDL}. These designs target different applications, including CPU cores~\cite{vexriscv, amid2020chipyard}, ML accelerators~\cite{amid2020chipyard}, vector arithmetic, cryptographic arithmetic, and other designs for logic synthesis study~\cite{brglez1989combinational}. 

\begin{table}[htbp]
      \centering
      \renewcommand{\arraystretch}{1.1}
      \resizebox{0.49\textwidth}{!}{
        \begin{tabular}{ |c||c|c|c|c| } 
        \hline
        Source      &    Number of    &  Original  &   Design Size (\#K Gates) \\  
        Benchmarks  &    Designs      &   HDL Type  &   \{Min, Median, Max\} \\  
        \hline
        \hline
        ISCAS'89~\cite{brglez1989combinational}$^\dagger$    & 5 & Verilog  & \{1, 6, 7\}  \\
        \hline
        ITC'99~\cite{corno2000rt}$^\dagger$     &  13 & VHDL   & \{4, 10, 45\}  \\
        \hline
        OpenCores~\cite{albrecht2005iwls}$^\dagger$  &  15 & Verilog &  \{2, 7, 62\} \\
        \hline 
        VexRiscv~\cite{vexriscv}   &  26 &  SpinalHDL & \{7, 132, 530\} \\
        \hline 
        RISC-V Cores$^\star$ &  5  & Verilog & \{7, 10, 17\} \\        
        \hline
        NVDLA~\cite{nvdla}  &  8  &  Verilog  & \{6, 40, 672\} \\
        \hline
        Chipyard~\cite{amid2020chipyard}$^\ddagger$ & 18 &  Chisel & \{1, 25, 921\}  \\
        \hline
        \end{tabular}
       }
        \begin{tablenotes}\scriptsize
        \item $^\dagger$ Tiny designs (i.e., $<$ 1K Gates) are removed from the original benchmarks.
        \item $^\star$ Collected open-source RISC-V cores~\cite{picorv32, mriscvcore} and their variants.
        \item $^\ddagger$ Rocket, BOOM and Sodor cores with  different configurations.
        \end{tablenotes}  
        \caption{Design RTL used for dataset generation. The comprehensive dataset encompasses diverse sources targeting various application scenarios, depicted in different HDL formats.
        } 
        \label{tbl:benchmark}
         %\vspace{-.2in}

\end{table}

For each design, the RTL is synthesized with Synopsys Design Compiler\textsuperscript{\textregistered} 2021 using the NanGate 45nm technology library~\cite{URL:NanGate}. The PPA values of the gate-level netlist of each design are recorded as the ground-truth label. Notice that we explored different synthesis parameters in Design Compiler\textsuperscript{\textregistered}. The label for each RTL design is determined based on the best PPA point on the Pareto curve, representing the best design trade-off designers may achieve. Actually, unlike described in some prior works~\cite{sengupta2022good}, the design trade-offs are not obvious using the latest commercial synthesis options, and we give a more detailed discussion in Subsection~\ref{sec:tradeoff}. 
The bit-level SOG representation is generated with open-source tools such as Yosys~\cite{wolf2013yosys} and Pyverilog~\cite{pyverilog}. 
The experiments are conducted on a server with a 2.9 GHz Intel Xeon(R) Platinum 8375C CPU, and 256G RAM. \looseness=-1

Based on the constructed dataset, the ML models can be implemented and evaluated. We adopt 10-fold cross-validation to assess the model accuracies. The hyper-parameter tuning for each ML model is based on a held-out validation set. After exploration and parameter tuning, our detailed model implementation is introduced below.

\subsubsection{Timing Models}
%Firstly, the linear function of each type of operator allows us to estimate the node delay for different fan-out numbers. Leveraging the node delay information, we map 20,000 RTL critical paths with their corresponding post-synthesis path slack labels. These critical paths serve as the training set for the path-level model, enabling us to establish a relationship between the RTL paths and netlist paths.
As mentioned in Section~\ref{sec:algorithm}, we explored two types of timing models for path delay prediction, as listed below. %The configurations of the models are listed below:

\begin{itemize}
\item Transformer: The transformer model shares the same hyperparameters as described in~\cite{xu2022sns} while we additionally introduce the fan-out of the operators into the path sequence.  
\item Random Forest: With the features listed above, we exploited the Random Forest model~\cite{louppe2014understanding} with 80 estimation trees and a maximum depth of 20 to perform the regression.
\end{itemize}

% \yao{Leave this to later results?} Since the Random Forest model, which incorporates human-extracted path features, demonstrates superior generalization and accuracy compared to the Transformer model, we select it as the preferred choice for the subsequent stages of our methodology.

% The evaluation results of two models are demonstrated in the first and second clusters of Figure~\ref{fig:bar}, where the Random Forest method based on the human-extracted path features shows better generalization and accuracy compared with the Transformer method. And therefore, we would employ this tree-based model in next stage.
% Since the Random Forest model based on careful human-extracted path features shows better generalization and accuracy compared with the Transformer method, we further utilized. 
% The Random Forest model based on careful human-extracted features is further utilized since it shows better accuracy compared with the Transformer method, demonstrated in Fig.~\ref{fig:bar}.
For the final design-level calibration model, we employed the XGBoost model with 45 estimators and a maximum depth of 8. %to calibrate the output estimated WNS and TNS from the path-level model, in conjunction with the features extracted from the RTL graph. 

\subsubsection{Power Models}
During the module-level power estimation, we explored and evaluated two different types of ML models:
\begin{itemize}
\item Graph Neural Network: we implemented a Graph Convolutional Network (GCN)~\cite{zhang2019graph} with two hidden convolutional layers with 10 and 70 nodes, respectively, and one sum-pooling layer. It performs an end-to-end graph-level value regression. We utilized the Adam optimizer with a learning rate 0.01 for model optimization, and the GCN model converges in 100 epochs.
\item XGBoost Regressor: With the features listed above, an XGBoost model~\cite{chen2016xgboost} with 30 estimators and a maximum depth of 6 is used for the regression task.
\end{itemize}

% The third cluster in Figure~\ref{fig:bar} illustrates the comparison between the GCN graph regression method and XGBoost Regressor combining the proposed features, where both the correlation coeffient R and the MAPE error indicate the tree-based method are much more capable of the module-level power estimation.

% Due to the XGBoost Regressor outperforms the GCN graph regression, we further utilize the tree-based method in terms of module-level power estimation.

% \yao{Leave this to later results?} As the XGBoost Regressor has shown superior performance in module-level power estimation, we utilize the output from the tree-based method for the next-stage design-level power calibration, which another tree-based ML model with the same configuration as used in the timing design-level modeling is exploited.
%Combining the sum of estimated module powers and SOG features, we exploit an additional tree-based method for the next-stage design-level power calibration, whose configuration is the same as used in the timing design-level model.

%The scatter plot in Fig.~\ref{fig:pwr} illustrates the power predictions of all designs, providing an overall view of the accuracy and effectiveness of the power modeling.
% limitation of suitable
An additional XGBoost model with 45 estimators and 8 max depths is utilized for design-level power calibration.

Due to the unavailability of power simulation testbenches for many designs, we demonstrate the vector-based power prediction capability of MasterRTL through the RISC-V CPU core series named BOOM~\cite{amid2020chipyard}, which can be simulated with widely-adopted Dhrystone testbench~\cite{weicker1984dhrystone} to obtain vector-based ground-truth power. All power models in this experiment are trained with vector-less power values. Unless explicitly indicated, the power prediction accuracy is also evaluated with vector-less power. 

%under specific workload scenario. 
% precise toggle rate and

% Based on the estimated power output from the module-level stage, we can conduct the last design-level power calibration, with the tree-based ML model that shares same configurations with that in the timing modeling method. The scatter plot of all designs is demonstrated in Figure~\ref{fig:pwr}.

\subsubsection{Area Model}
Similar to the design-level timing and power model, the XGBoost regressor with 45 estimators and 12 max depth is evaluated for the combinational area prediction. The total area is predicted by adding the predicted combinational area with the directly calculated sequential area. %The detailed scatter plot is shown in Fig.~\ref{fig:area}.

We use three metrics to evaluate the prediction accuracy between the predicted value $\hat{y}$ and ground-truth $y$ of $n=90$ designs. They are: correlation coefficient (R), Mean Absolute Error Percentage (MAPE), and Root Relative Square Error (RRSE), as defined below.
\begin{equation*}
\text{MAPE}= \dfrac{1}{n}\sum_{i=1}^{n}\dfrac{|y_i-\hat{y_i}|}{y_i} \times 100\%\footnote{To mitigate the influence of a few extreme outliers, MAPE values exceeding 100\% are capped at 100\% to prevent disproportionate impact on the metric.}, \ \ \text{RRSE} = \sqrt{\dfrac{\sum_{i=1}^{n}{(y_i-\bar{\hat{y_i}}})^2}{\sum_{i=1}^{n}{(y_i-\bar{y}})^2}}
\end{equation*}
The $\bar{\hat{y}}$ and $\bar{y}$ represent the average of predicted and measured values, respectively. These metrics bring a comprehensive and fair evaluation of ML models from three aspects, where the higher correlation R and lower MAPE and RRSE indicate better accuracy. %Specifically, R measures a linear correlation between predicted values and the ground truth values, MAPE gives an intuitive percentage of disparity, while RRSE can represent the accuracy ignoring the influence of different RTL design scales.
% Note that we set the MAPE of the prediction on one design as 100\% whenever it is larger than 100\%, so that the huge influence of a small amount of outliers can be eliminated.

\begin{figure}[!h]
  \centering
  \includegraphics[width=0.8\linewidth]{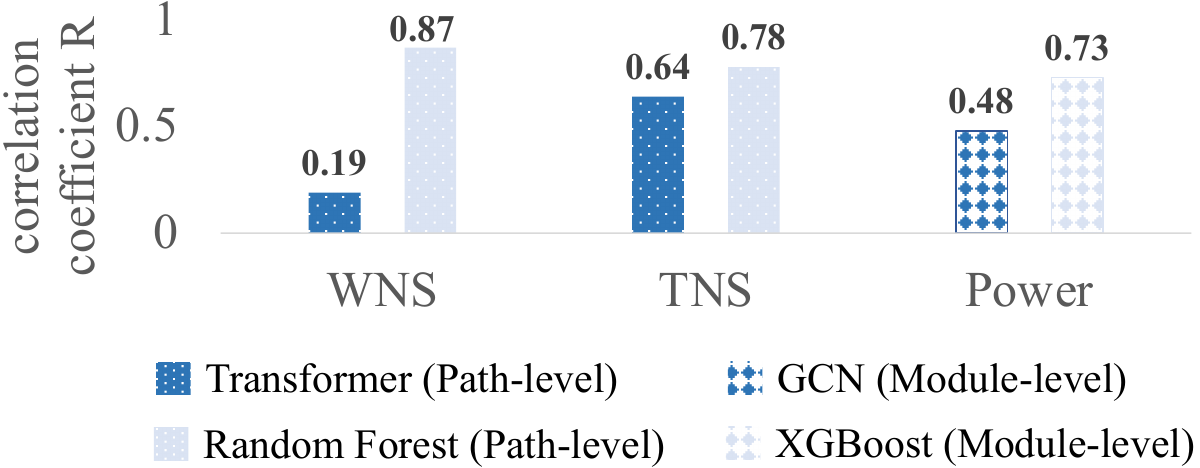}
  \caption{%Correlation Coefficient R and Mean Absolute Error Percentage (MAPE) of ML models used for timing, power, and area prediction at different stages.  XGBoost Regressor is employed for calibrating PPA estimations in each final design-level stage, contributing to the high accuracy predictions.
  Evaluation of intermediate-stage ML models in MasterRTL. 
  Tree-based models are more accurate in path-level timing and module-level power modeling, outperforming deep learning models (i.e., Transformer and GCN).}
  \label{fig:bar}
\end{figure}

% accuracy of 
Fig~\ref{fig:bar} compares these explored intermediate-stage ML models in MasterRTL, and the final prediction results will be introduced later in Subsection~\ref{sec:results}. 
For path-level timing models, the Random Forest model with human-extracted path features performs better than Transformer. The trend is similar in the module-level models, where the XGBoost regressor is more accurate than the GCN model. Therefore, the traditional tree-based model is finally adopted at the intermediate stages of MasterRTL instead of deep learning models. 

% to provide essential features for the design-level models 

\iffalse
\begin{table}[!t]
      \centering
      \renewcommand{\arraystretch}{1.1}
      \resizebox{0.47\textwidth}{!}{
        \begin{tabular}{ |c|c|c|c|c| } 
        \hline
        \multirow{2}{*}{Works} &    RTL Design       &   \multirow{2}{*}{Primarily Modeling} \\  
                               &    Representation   &    \\  
        \hline
        \hline
        ICCAD'22~\cite{sengupta2022good}  & Word-level & All register trees \\
        \hline
        ISCA'22~\cite{xu2022sns}    &  Word-level  & Randomly sampled paths \\
        \hline
        \hline
        This work      &    Bit-level & Longest paths, all modules \\
        \hline
        \end{tabular}
       }
        \caption{Methodolodies of RTL-stage cross-design PPA modeling solutions.} 
        \label{tbl:related2}
         \vspace{-.2in}
\end{table}
\fi

\subsection{Baseline Solutions and their Limitations}
\label{sec:baseline}

As mentioned, \cite{xu2022sns, sengupta2022good} are obviously state-of-the-art cross-design RTL modeling methods, which should be compared with our solution. The implementation of~\cite{xu2022sns} is open-sourced. Although \cite{sengupta2022good} is not open-sourced, by strictly following their description, we can implement their solution without much doubt. Neither works~\cite{xu2022sns, sengupta2022good} released their dataset. For a fair comparison, both baselines and MasterRTL are trained and evaluated on exactly the same training and testing data, with 10-fold cross-validation (i.e., each time 90\% designs for training, 10\% for testing, no design trained and tested at the same time). Before we delve into the experimental results, we will first inspect these solutions~\cite{xu2022sns, sengupta2022good} in detail, and discuss their potential limitations, which lead to inferior performance.

\begin{figure}[!h]
	\centering
        \subfloat[Influence of overlaps in~\cite{sengupta2022good}]{\includegraphics[width=0.46\linewidth]{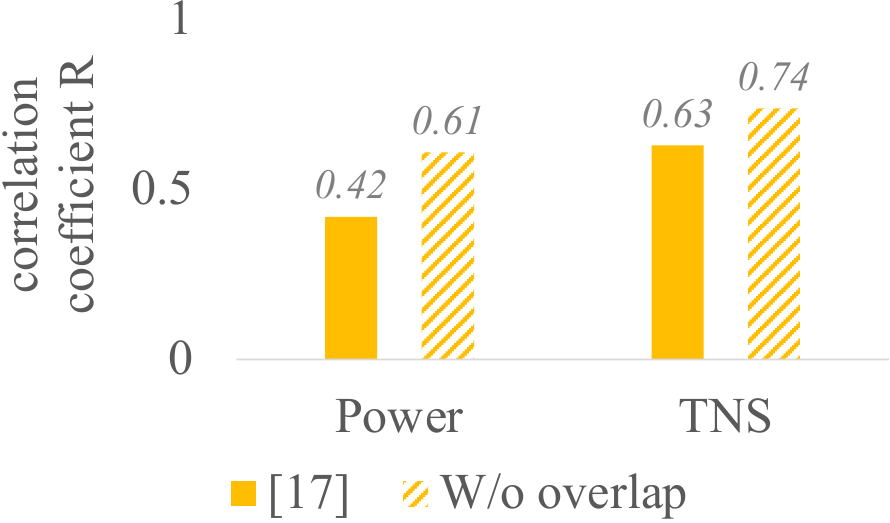}\label{fig:overlap}}\hspace{5pt}
	% \subfloat[Distinct distribution between training and testing paths in RTL designs in~\cite{xu2022sns}.]
        \subfloat[Slack distribution gaps in~\cite{xu2022sns}]{\includegraphics[width=0.46\linewidth]{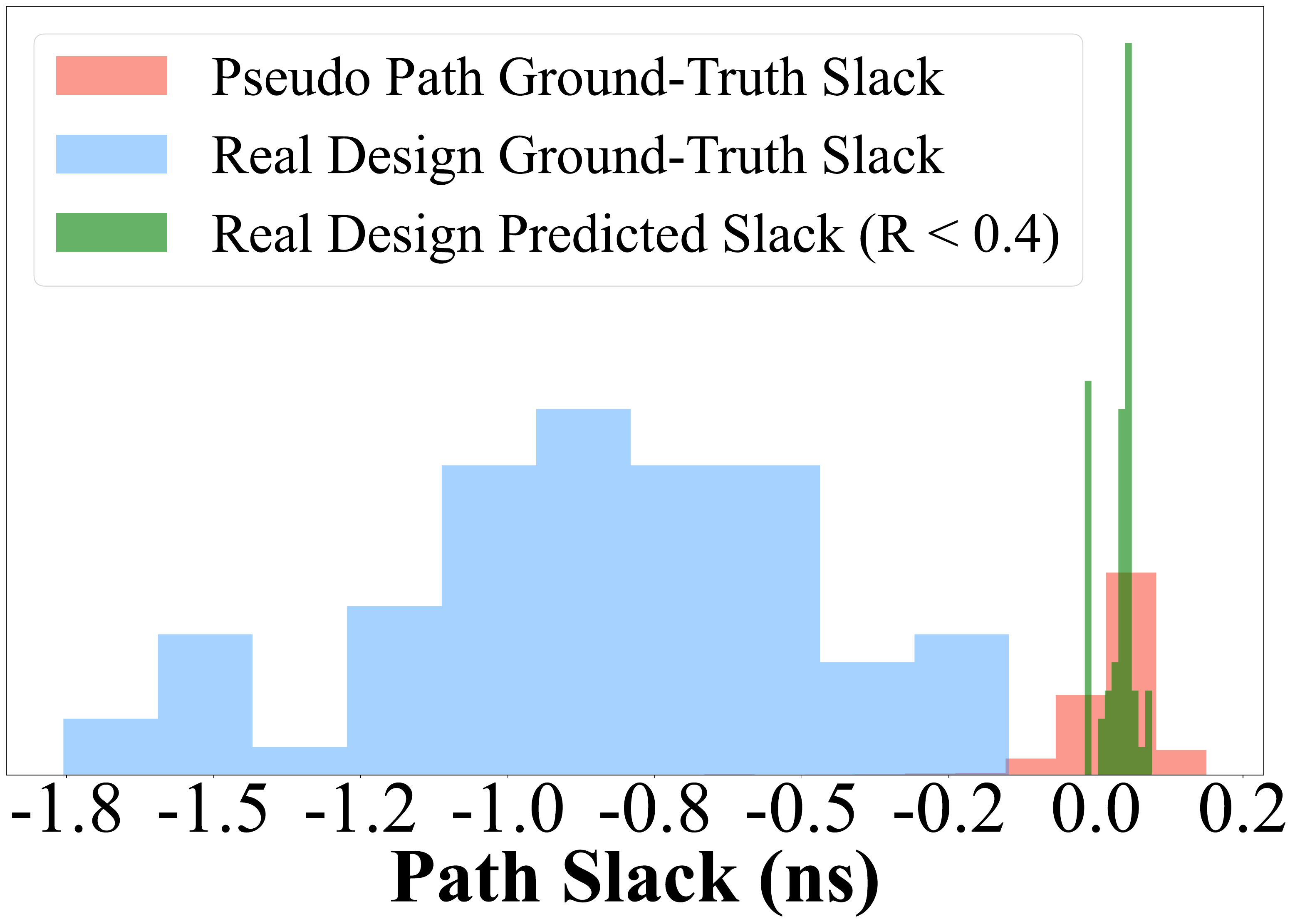}\label{fig:sns}}\hspace{0pt}
        \caption{Limitations of prior works. (a) Overlapping logic among different register trees in~\cite{sengupta2022good} negatively affects accuracy. %, improved by removing overlaps. 
        (b) In~\cite{xu2022sns}, the distinct distributions between training paths and target designs result in inaccuracies when predicting path slack.}
        \label{fig:limit}
\end{figure}

The work of~\cite{sengupta2022good} first backtraces every register to build the logic tree driving this register, named a register tree. After that, it calculates pre-defined features for each register tree and sums up the features of all register trees, generating the final ML model inputs. However, we observe significant logic overlaps among different register trees, resulting in the same RTL logic being counted multiple times as undesired duplications.  
As our experiment in Fig.~\ref{fig:limit}(a) shows, the accuracies on power and TNS both increase when such an overlap is removed.\looseness=-1

\begin{table*}[!t]
\centering
\renewcommand{\arraystretch}{1.05}
\resizebox{0.99\textwidth}{!}{
%\begin{tabular}{|c|c||c|c|c||c||c|c|c|c||c|c|c|c||c|c|c|}
\begin{tabular}{|c||c|c|c|c||c|c|c|c||c|c|c|c||c|c|c|c|}
\hline
\multirow{2}{*}{Method}  & \multirow{2}{*}{Target} & \multirow{2}{*}{R}    & \multirow{2}{*}{MAPE} & \multirow{2}{*}{RRSE} & \multirow{2}{*}{Target} & \multirow{2}{*}{R}    & \multirow{2}{*}{MAPE} & \multirow{2}{*}{RRSE} & \multirow{2}{*}{Target} & \multirow{2}{*}{R}    & \multirow{2}{*}{MAPE} & \multirow{2}{*}{RRSE} & \multirow{2}{*}{Target} & \multirow{2}{*}{R}    & \multirow{2}{*}{MAPE} & \multirow{2}{*}{RRSE} \\ & & & & & & & & & & & & & & & &  \\ 
\hline
\multicolumn{10}{c}{}\\[-0.8em]
\hline
\cellcolor{orangeShallow}\textbf{\cite{sengupta2022good} (AST-alike)}          & \multirow{6}{*}{WNS}        & 0.37 & 26\%        & 0.95 & \multirow{6}{*}{TNS}        & \cellcolor{orangeShallow}\textbf{0.63} & \cellcolor{orangeShallow}\textbf{48\%} & \cellcolor{orangeShallow}\textbf{0.79} & \multirow{6}{*}{\begin{tabular}[c]{@{}c@{}}Total \\      Power\end{tabular}} & \cellcolor{orangeShallow}\textbf{0.42} & \cellcolor{orangeShallow}\textbf{51\%} & \cellcolor{orangeShallow}\textbf{1.01} & \multirow{6}{*}{Area}       & 0.75 & 38\%        & 0.68 \\ \cline{1-1} \cline{3-5} \cline{7-9} \cline{11-13} \cline{15-17} 
\cite{sengupta2022good} (SOG)                &                             & 0.82 & 24\%        & 0.53 &                             & 0.86 & 41\%        & 0.36 &                                                                              & 0.62 & 48\%        & 0.79 &                             & 0.94 & 31\%        & 0.33 \\ 
% \cline{1-1} \cline{3-5} \cline{7-9} \cline{11-13} \cline{15-17} 

\hhline{-~---~---~---~---}\noalign{\vskip-2\tabcolsep \vskip-3\arrayrulewidth \vskip\doublerulesep}\\
&  \multicolumn{10}{c}{}\\[-1.2em]
\hhline{-~---~---~---~---}

\cellcolor{lightgreen}\textbf{\cite{xu2022sns} (AST-alike)}          &                             & \cellcolor{lightgreen}\textbf{0.71} & \cellcolor{lightgreen}\textbf{35\%}        & \cellcolor{lightgreen}\textbf{1.15} &                             & 0.78 & 36\%        & 1.1  &                                                                              & \cellcolor{lightgreen}\textbf{0.74} & \cellcolor{lightgreen}\textbf{68\%}        & \cellcolor{lightgreen}\textbf{0.81} &                             & \cellcolor{lightgreen}\textbf{0.93} & \cellcolor{lightgreen}\textbf{38\%} & \cellcolor{lightgreen}\textbf{0.42} \\ \cline{1-1} \cline{3-5} \cline{7-9} \cline{11-13} \cline{15-17} 
\cite{xu2022sns} (SOG)                &                             & 0.78 & 26\%        & 0.78 &                             & 0.8  & 29\%        & 0.88 &                                                                              & 0.82 & 48\%        & 0.62 &                             & 0.96 & 35\%        & 0.4  \\
% \cline{1-1} \cline{3-5} \cline{7-9} \cline{11-13} \cline{15-17} 
\hhline{-~---~---~---~---}\noalign{\vskip-2\tabcolsep \vskip-3\arrayrulewidth \vskip\doublerulesep}\\
&  \multicolumn{10}{c}{}\\[-1.2em]
\hhline{-~---~---~---~---}
MasterRTL   (AST-alike) &                             & 0.81 & 22\%        & 0.6  &                             & 0.95 & 36\%        & 0.31 &                                                                              & 0.79 & 44\%        & 0.63 &                             & 0.94 & 31\%        & 0.34 \\ \cline{1-1} \cline{3-5} 
% \cline{7-9} \cline{11-13} \cline{15-17} 
\hhline{-~---~---~---~---}
\cellcolor{cyan}\textbf{MasterRTL (SOG)}         &                             & \cellcolor{cyan}\textbf{0.93} & \cellcolor{cyan}\textbf{14\%} & \cellcolor{cyan}\textbf{0.4}  &  & \cellcolor{cyan}\textbf{0.96} & \cellcolor{cyan}\textbf{27\%} & \cellcolor{cyan}\textbf{0.29} &  & \cellcolor{cyan}\textbf{0.89} & \cellcolor{cyan}\textbf{38\%}        & \cellcolor{cyan}\textbf{0.54} &                             & \cellcolor{cyan}\textbf{0.98} & \cellcolor{cyan}\textbf{16\%}        & \cellcolor{cyan}\textbf{0.24} \\ \hline
\end{tabular}}
\caption{Accuracy comparison for WNS, TNS, total power and area evaluations. Colored rows represent originally proposed methods. The prior work ICCAD'22~\cite{sengupta2022good} only proposes to evaluate TNS and power, and ISCA'22~\cite{xu2022sns} proposes to evaluate WNS, power and area.}
\label{tbl:timing_accuracy}
\vspace{-.1in}
\end{table*}

\begin{figure*}[!t]
\captionsetup[subfigure]{labelformat=empty}
	\centering
	\subfloat[]{\includegraphics[height=0.22\linewidth, max width=0.242\linewidth]{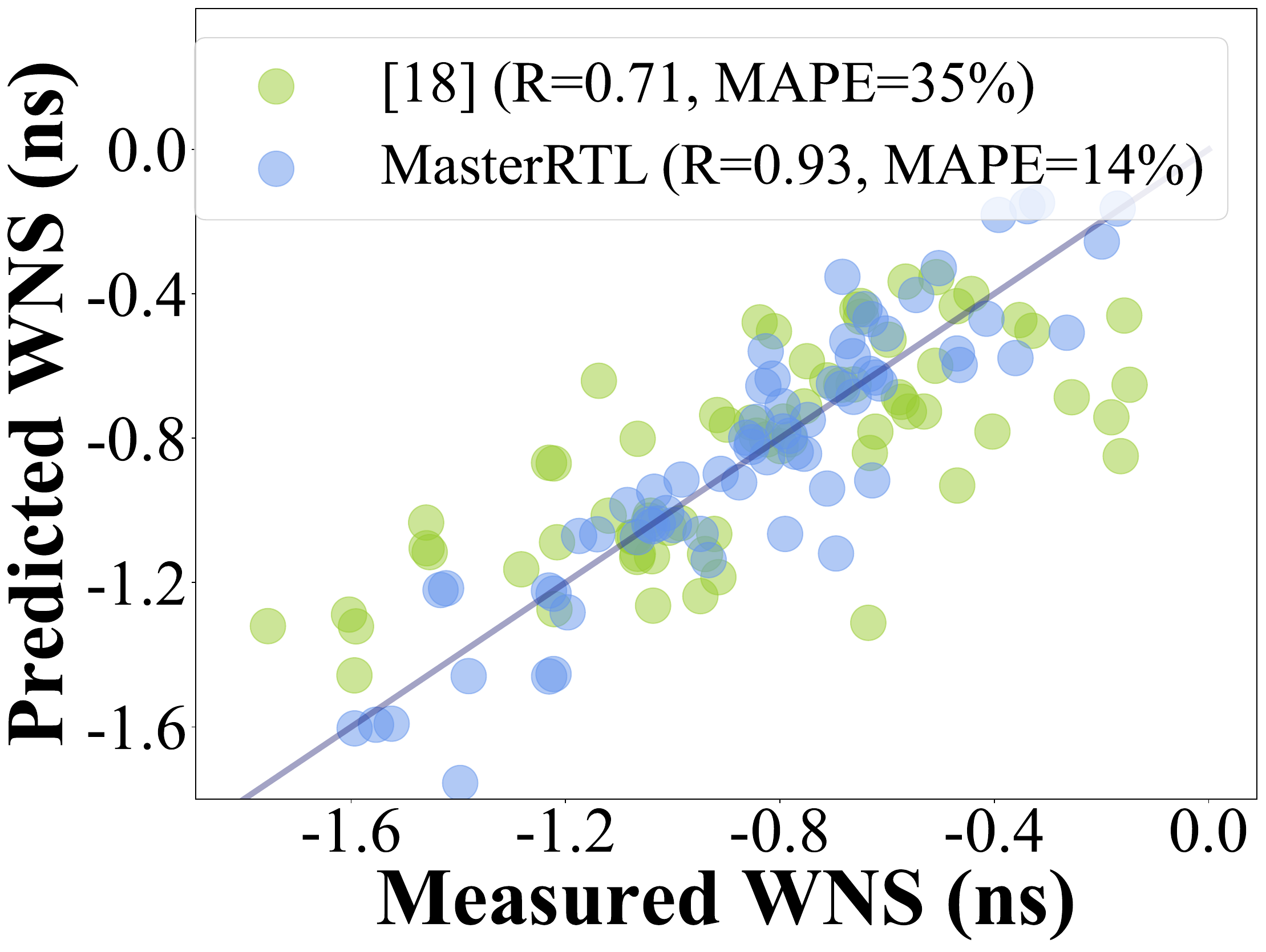}}\hspace{0pt}
	\subfloat[]{\includegraphics[height=0.22\linewidth, max width=0.242\linewidth]{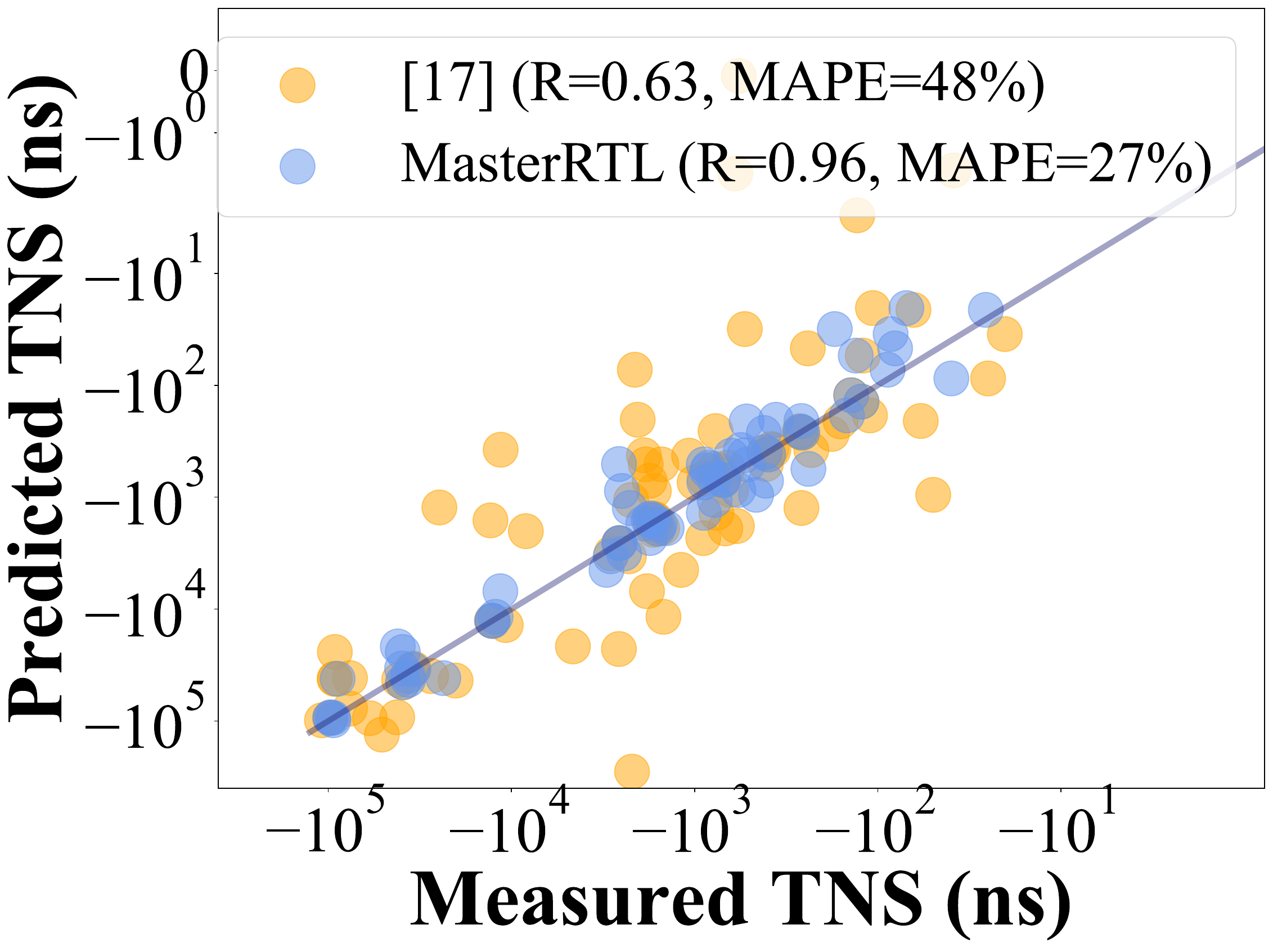}}\hspace{0pt}
	\subfloat[]{\includegraphics[height=0.22\linewidth, max width=0.242\linewidth]{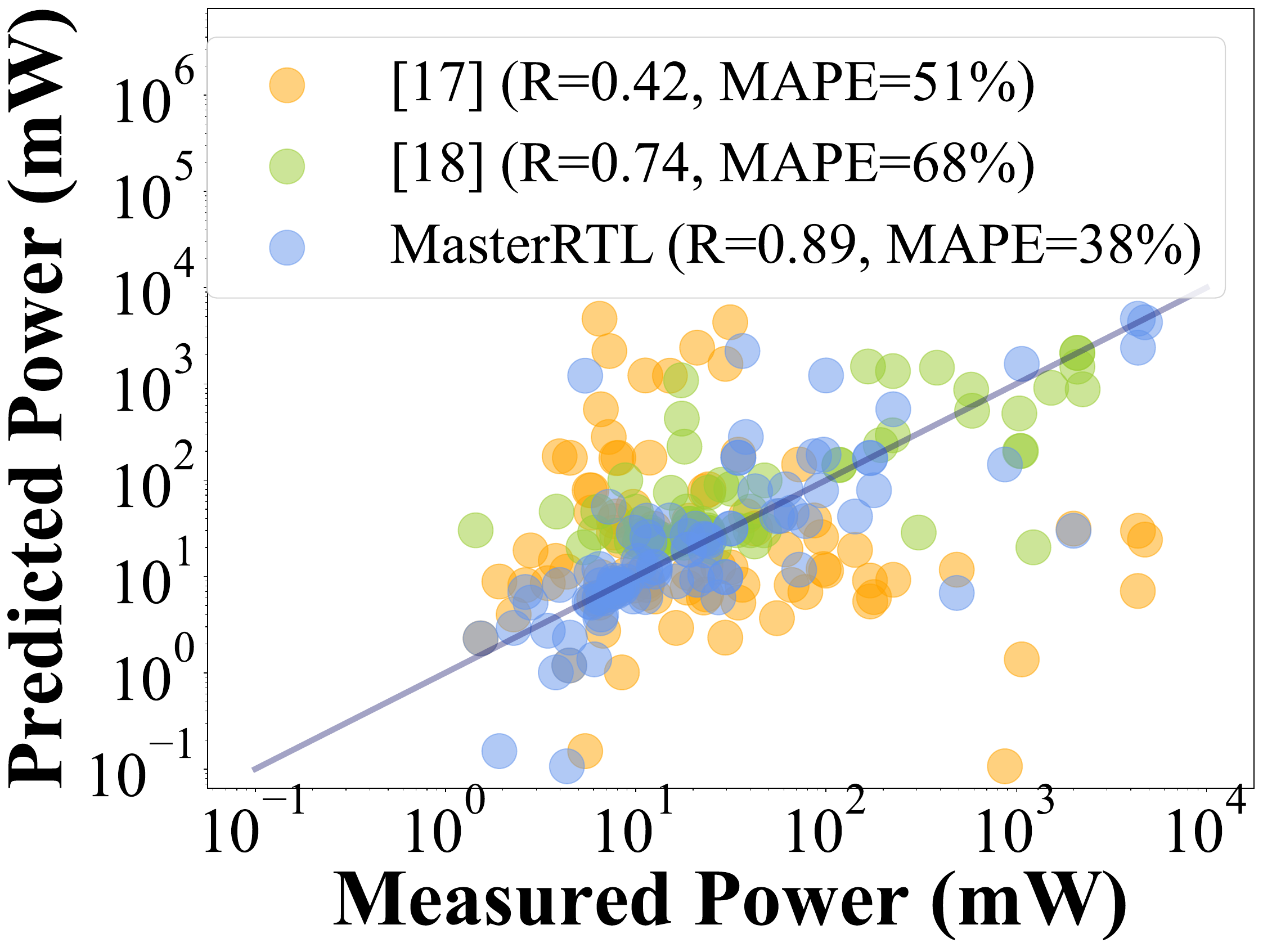}}\hspace{0pt}
	\subfloat[]{\includegraphics[height=0.22\linewidth, max width=0.242\linewidth]{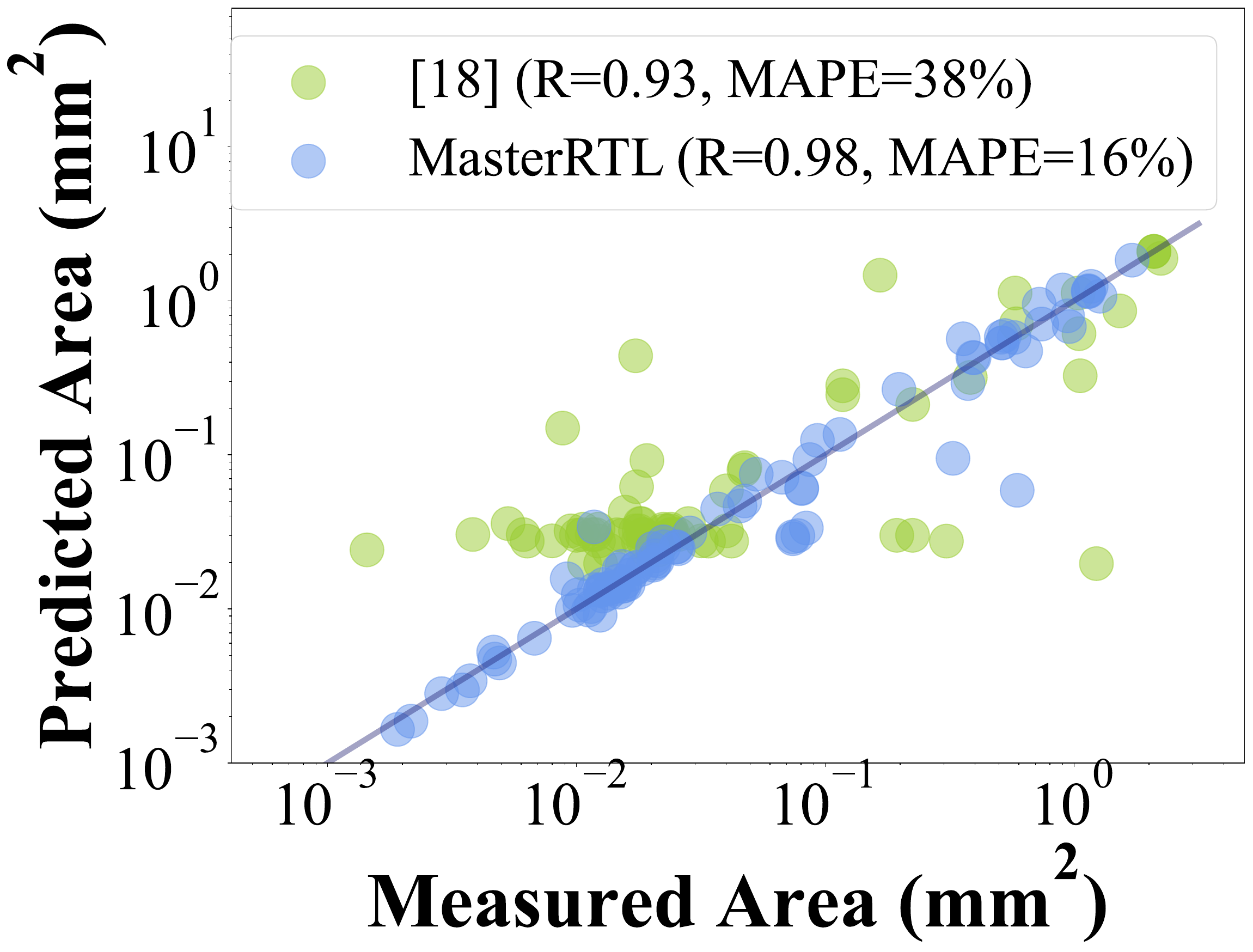}}\hspace{0pt}
 \vspace{-.2in}
	\caption{
    % MasterRTL prediction accuracy for all PPA characteristics of each Design. 
    Prediction vs ground-truth for each PPA characteristic of all designs.
    Comparing with the SOTA solutions~\cite{sengupta2022good, xu2022sns},
    % The scatter points for MasterRTL and the baselines~\cite{sengupta2022good, xu2022sns} are presented in blue, orange, and green scattered points, respectively. 
    MasterRTL significantly outperforms prior works in all PPA predictions.}
        % \caption{Prediction vs ground truth for}
        \label{fig:scatter}
        \vspace{-.1in}
\end{figure*}

The work of~\cite{xu2022sns} trains ML models to predict the PPA of each individual path. To generate the training dataset, it generates pseudo individual RTL paths, and then synthesizes them to collect the PPA label of each pseudo path. After training, the ML model is applied to infer paths sampled randomly from the target design. However, as shown in Fig.~\ref{fig:limit}(b), we observe a significant gap between the pseudo training paths and real test paths from realistic target designs. These distinct distributions imply that the slack predictions (green) are misled by the pseudo paths (red), resulting in a large gap between predictions and real test paths (blue) in both absolute error and correlation R. In addition, the randomly sampled paths cannot fully reflect the TNS and WNS of the target design. 

%Our experiments show that .. 

In the experimental results, we will emphasize and primarily compare with their originally proposed methods~\cite{xu2022sns, sengupta2022good}. Besides that, we have also implemented many baseline variations for detailed ablation studies. For all setups and variations, we have tried to maximize the performance through the similar model-tuning procedure.

% \begin{figure}
%   \centering
%   \includegraphics[width=0.49\linewidth]{_fig/transformer.pdf}

%   \caption{Trade off}
%   \label{fig:tradeoff}
% \end{figure}

\begin{table}[!b]
      \centering
      \renewcommand{\arraystretch}{1.1}
      \resizebox{0.49\textwidth}{!}{
            \begin{tabular}{|c||cccc|}
\hline
\multirow{3}{*}{\textbf{Design}}  & \multicolumn{4}{c|}{\textbf{MAPE (\%)}}                                                                          \\ \cline{2-5} 
                         & \multicolumn{3}{c|}{\textbf{Vector-less Method}  }                                          & \textbf{Vector-based Method} \\ \cline{2-5} 
                         & \multicolumn{1}{c|}{~\cite{ sengupta2022good}} & \multicolumn{1}{c|}{~\cite{xu2022sns}} & \multicolumn{1}{c|}{MasterRTL} & MasterRTL          \\ \hline \hline
SmallBOOM                & \multicolumn{1}{c|}{84} & \multicolumn{1}{c|}{59} & \multicolumn{1}{c|}{42}        & 34                 \\ \hline
MediumBOOM               & \multicolumn{1}{c|}{74} & \multicolumn{1}{c|}{50} & \multicolumn{1}{c|}{21}        & 5                  \\ \hline
LargeBOOM                & \multicolumn{1}{c|}{67} & \multicolumn{1}{c|}{58} & \multicolumn{1}{c|}{39}        & 8                  \\ \hline \hline
Aver. MAPE (BOOM)        & \multicolumn{1}{c|}{75} & \multicolumn{1}{c|}{56} & \multicolumn{1}{c|}{34}        & 16                 \\ \hline
Aver. MAPE (All designs) & \multicolumn{1}{c|}{51} & \multicolumn{1}{c|}{68} & \multicolumn{1}{c|}{38}        & N/A                  \\ \hline
\end{tabular}
    }
    \caption{Comparison between the MasterRTL vector-less and vector  power prediction results. Trained on vector-less power values, MasterRTL performs well for vector-based power prediction.}%The accuracy is further enhanced by introducing the power simulation.}  
        \label{tbl:vecpwr}
\end{table}

\subsection{PPA Estimation Accuracy Evaluation and Comparison}
\label{sec:results}

% \yao{Better first compare with baselines, leave Fig.8 later. Fig.8 is less important.}

Table~\ref{tbl:timing_accuracy} shows  the comparison of MasterRTL over prior works~\cite{xu2022sns, sengupta2022good}. Notice that only the colored cells with bold texts denote actual estimation methods, while other cells are implemented for ablation studies. As we have explained, prior works adopt AST-alike representations while MasterRTL uses the SOG. \cite{sengupta2022good} originally only predicted TNS and power, while \cite{xu2022sns} was for WNS, power and area.\looseness=-1

%We have multiple interesting observations in Table~\ref{tbl:timing_accuracy}. First, MasterRTL significantly outperforms original prior works for both TNS ($R=0.96>0.65$~\cite{sengupta2022good}) and WNS ($R=0.93 > 0.71$~\cite{xu2022sns}) estimation. MasterRTL can achieve $R>0.90$ for both TNS and WNS. Second, for all methods, the advantage of our proposed bit-level representation over the word-level representation is not limited to MasterRTL. All methods using bit-level representation generally perform better. The advantage is more obvious for~\cite{sengupta2022good}, while less obvious for~\cite{xu2022sns}. Third, the accuracy of .  

We have multiple interesting observations in Table~\ref{tbl:timing_accuracy}. First of all, MasterRTL significantly outperforms original prior works for all originally proposed estimations, specifically, WNS ($R=0.93 > 0.71$~\cite{xu2022sns}), TNS ($R=0.96>0.63$~\cite{sengupta2022good}), power ($R=0.89 > 0.42$~\cite{sengupta2022good} and $0.74$~\cite{xu2022sns}), area ($R=0.98 > 0.93$~\cite{xu2022sns}), and also much lower MAPE and RRSE errors. Another key observation is that the advantage of our proposed SOG over AST-alike representation is not limited to MasterRTL. All methods using SOG generally perform better. The advantage is more obvious for~\cite{sengupta2022good}, while less obvious for~\cite{xu2022sns}. This universal accuracy improvement validates our claimed advantages of SOG, which we believe should be more widely adopted as a highly ML-friendly RTL representation. 

% In addition,  the advantage of SOG is more obvious for challenging tasks that estimate  WNS and power,  while TNS and area are more correlated with the scale of each design and are therefore, easier to predict. Scatter plots in Fig.~\ref{fig:scatter} detail the comparisons of the colored cells in Table~\ref{tbl:timing_accuracy}.
In addition, compared with the TNS and area, which are more correlated with each design's scale, the WNS and power estimations are generally more challenging and less accurate. The advantage of SOG over AST-alike is also more obvious for these challenging tasks. More detailed comparisons of the colored cells in Table~\ref{tbl:timing_accuracy} are shown by the scatter plots in Fig.~\ref{fig:scatter}.

Table~\ref{tbl:vecpwr} further shows the vector-based power prediction on BOOM series CPU designs, which support simulation based on testbenches like Dhrystone. Table~\ref{tbl:vecpwr} indicates by adopting toggle rate in power model features, MasterRTL trained with vector-less power values only can directly predict vector-based power accurately. This power model unifies both vector-less and vector-based power analysis scenarios. \looseness=-1

%the accuracy improvement when the simulated toggle rate is introduced.
%results with the vector-less ones

\subsection{PPA Estimation Accuracy Ablation Study}

To analyze the superior performance of MasterRTL, we decompose our solution to provide ablation studies by removing key policies of MasterRTL. The following summarizes the crucial policies in MasterRTL: 1) SOG rather than AST-alike representation; 2) customized key features for WNS, TNS, and power, respectively (e.g. critical path information from path-level model for timing, the introduction of toggle rate and module-level partition for power); 3) design-level calibration.

\begin{figure}[h]
  \centering
  \includegraphics[width=1\linewidth]{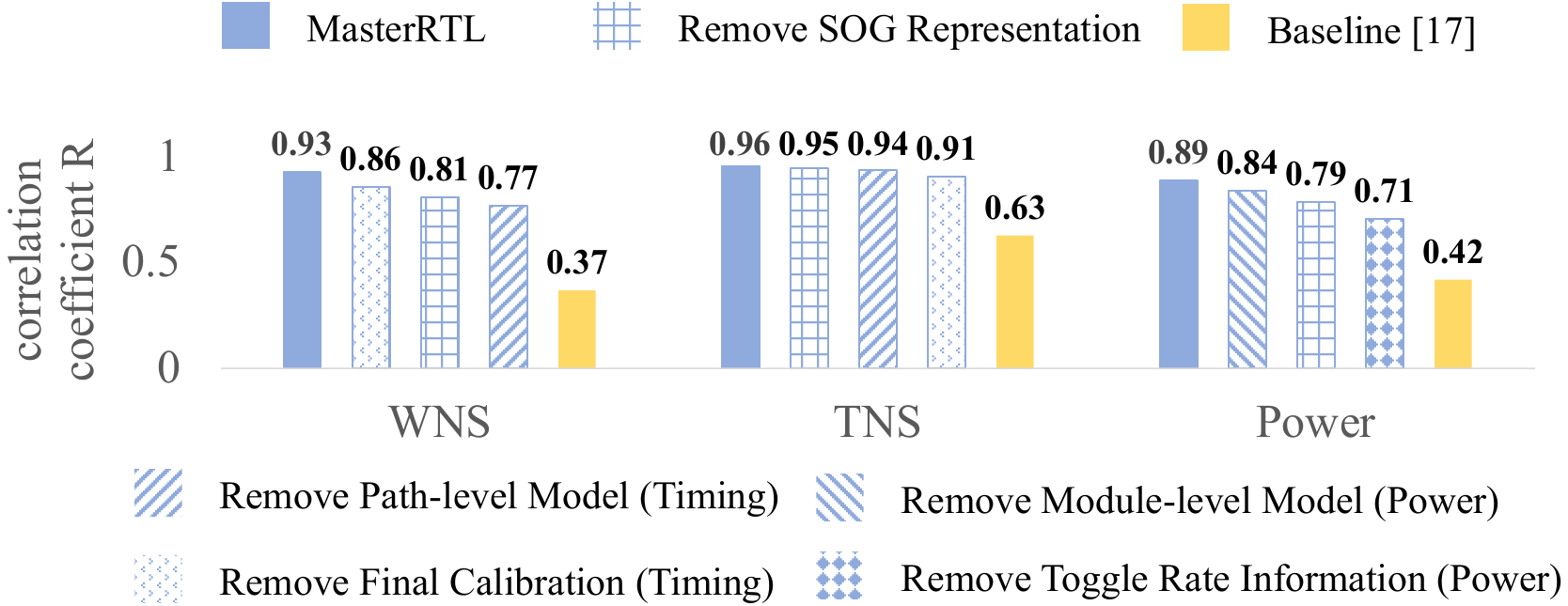}
  \caption{Ablation study of decomposed factors contributing to the PPA estimation accuracy. Except for the SOG representation used in all models, the inclusion of customized key features is crucial for accuracy enhancement.}
  \label{fig:abla}
  %\vspace{-.1in}
\end{figure}

\begin{figure*}[!t]
\captionsetup[subfigure]{labelformat=empty}
\vspace{-.1in}
	\centering
	\subfloat[]{\includegraphics[height=0.2\linewidth, max width=0.242\linewidth]{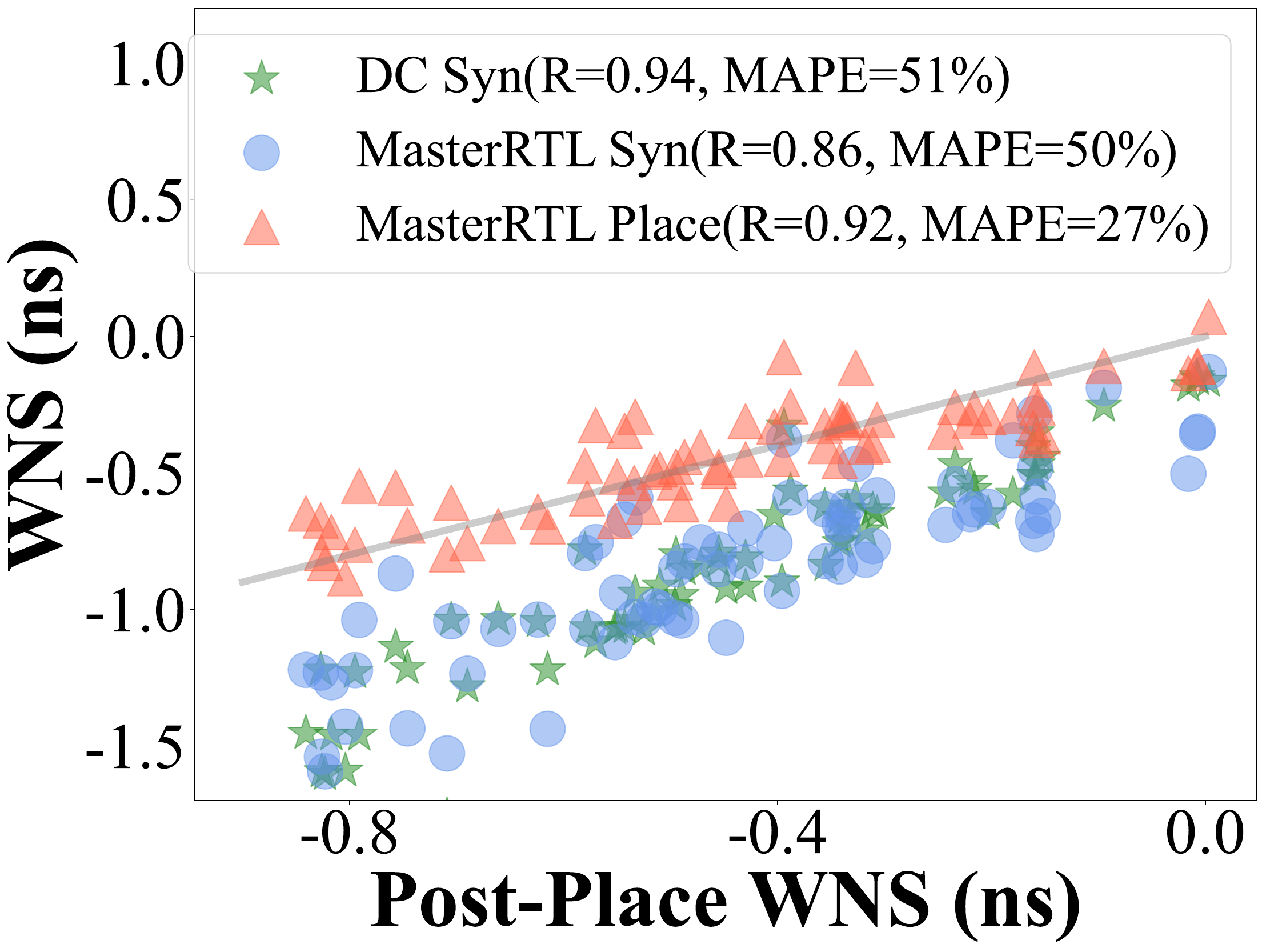}\label{fig:wns}}\hspace{0pt}
	\subfloat[]{\includegraphics[height=0.2\linewidth, max width=0.242\linewidth]{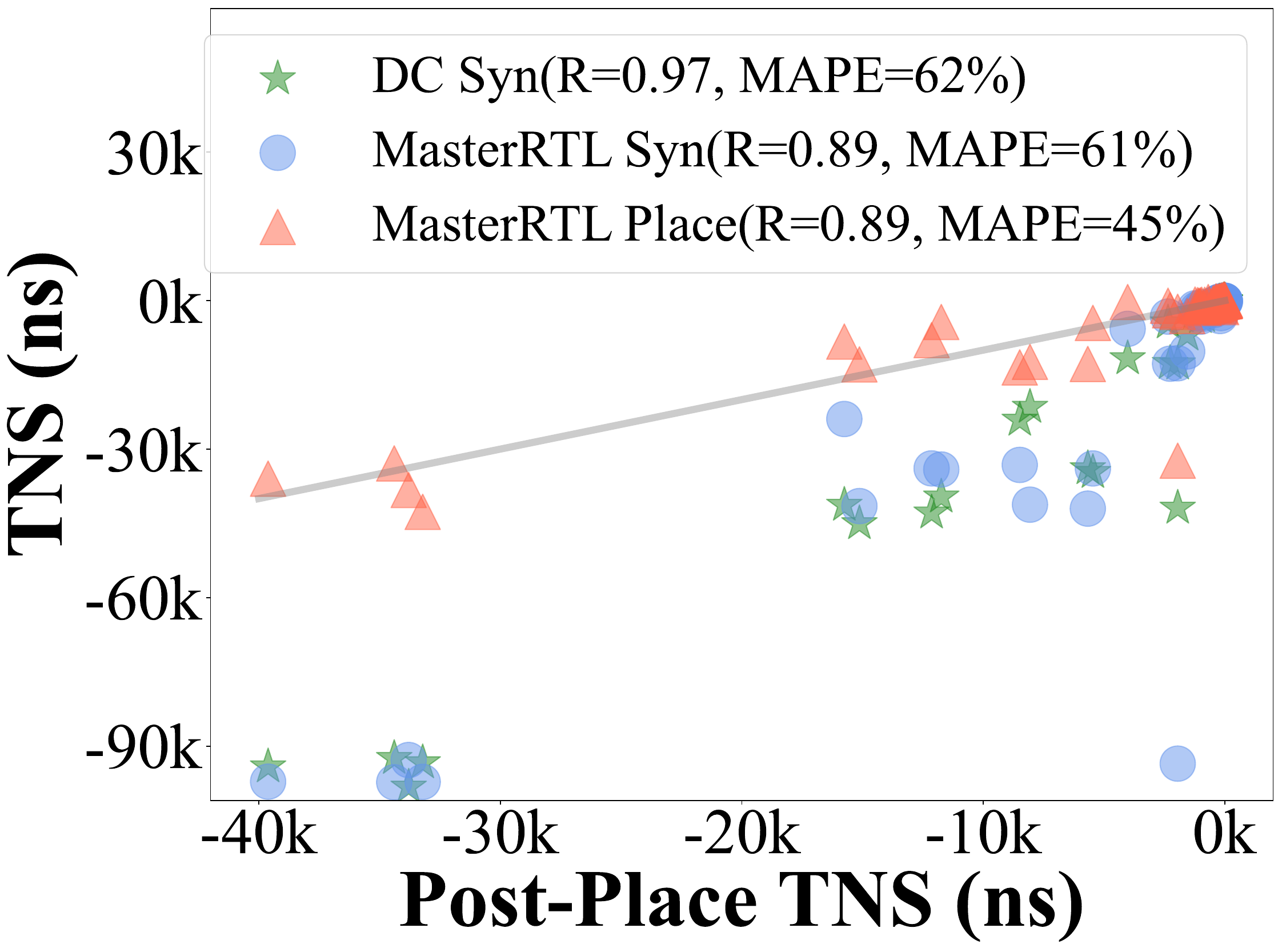}\label{fig:tns}}\hspace{0pt}
	\subfloat[]{\includegraphics[height=0.2\linewidth, max width=0.242\linewidth]{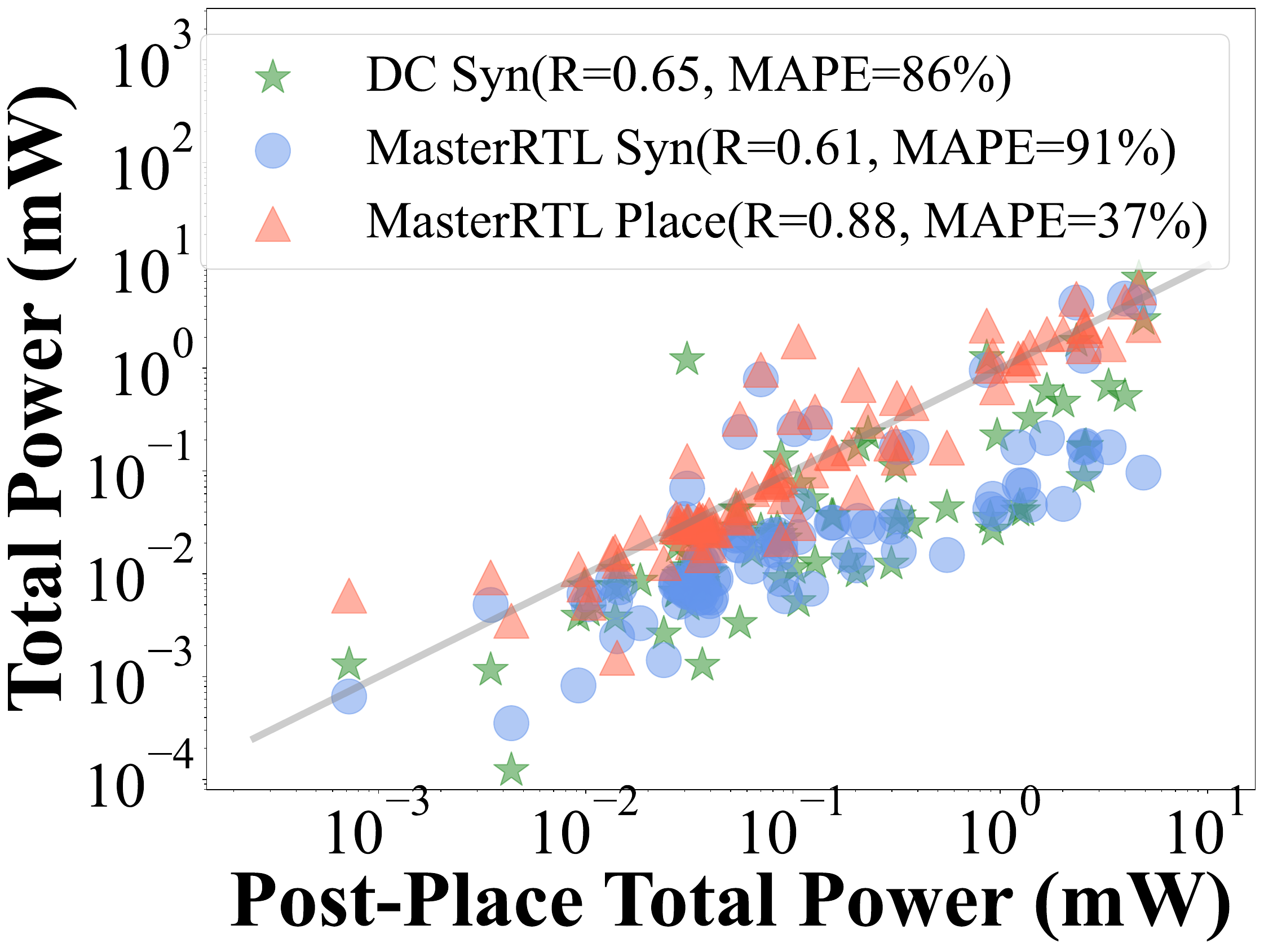}\label{fig:pwr}}\hspace{0pt}
	\subfloat[]{\includegraphics[height=0.2\linewidth, max width=0.242\linewidth]{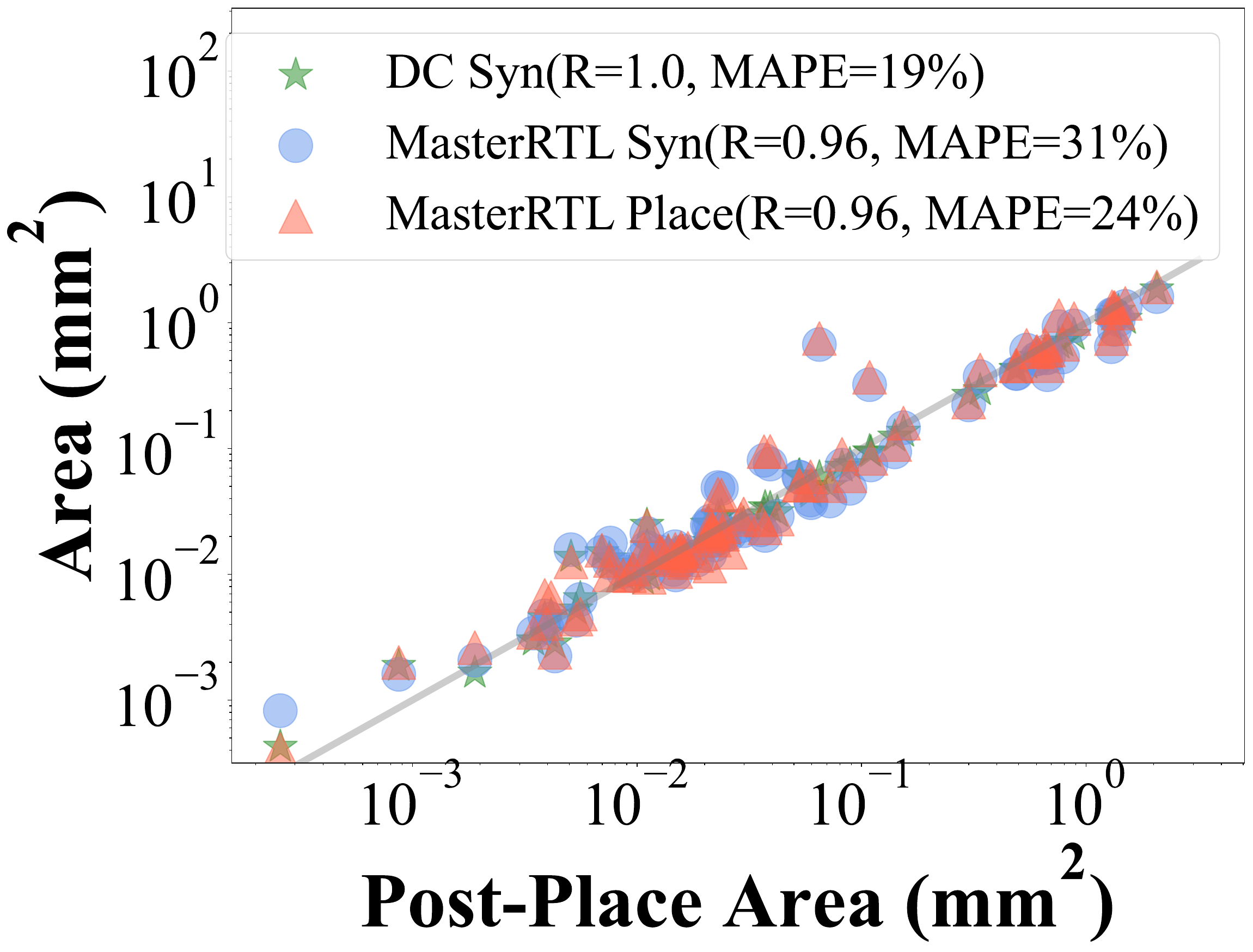}\label{fig:area}}\hspace{0pt}
 \vspace{-.2in}
	\caption{Extending MasterRTL for placement PPA prediction. The post-synthesis netlist's ground-truth and predictions are in green and blue points, while the red ones are the post-placement PPA prediction from extended MasterRTL. After being extended to placement PPA prediction, MasterRTL achieves similar or even higher accuracy on placement solutions than ground-truth netlist. Notice that it still makes predictions at RTL-stage.}
        \label{fig:layout}
          \vspace{-.1in}
\end{figure*}

Fig.~\ref{fig:abla} shows ablation study results. For WNS prediction, which heavily relies on critical path information, removing the path-level model leads to the most substantial accuracy drop. Regarding TNS, the most significant accuracy drop happens when graph features are removed, indicating the importance of SOG features and design scale. For the total power, the toggle rate information is crucial to the accuracy.
%which indicates they are mainly influenced by relationships captured from SOG as well as the entire design scale. 
% Finally, area prediction is significantly affected by the analytical area, which is calculated also based on the graph features.
After removing all the proposed crucial policies, the final accuracy is similar to the baseline method~\cite{sengupta2022good}.

\subsection{MasterRTL Runtime Overhead}
MasterRTL, functioning as an RTL-stage estimator, offers accurate PPA evaluation without a time-consuming logic synthesis process. To evaluate its efficiency, we present the median runtime of all 90 benchmark designs. Table~\ref{tbl:runtime} compares the runtime overhead of MasterRTL with the commercial logic synthesis tool.
Our framework exhibits a runtime overhead of approximately 5\% compared to the synthesis runtime. The primary contributor to this overhead is the preprocessing of RTL designs, particularly the conversion from RTL designs to SOG, which consumes 3.4\% of the synthesis runtime. Additionally, the extraction of toggle rate in synthesis tools accounts for 0.8\% of the synthesis time. The feature engineering processes, on the other hand, typically require only seconds to extract features by traversing the entire design SOG, accounting for no more than 0.5\% of the synthesis time. Lastly, the inference of all PPA values through our ML-based timing, power, and area models requires less than 0.1 seconds.\looseness=-1

\begin{table}[h]
  \centering
  \renewcommand{\arraystretch}{1.1}
  \resizebox{0.45\textwidth}{!}{
    \begin{tabular}{|c|c||c|c|c|}
    \hline
    \multicolumn{2}{|c||}{\textbf{Stage}}   & \textbf{Runtime$^\star$}      \\
    \hline
    \hline
    \multirow{2}{*}{RTL Preprocess} & SOG Construction & 3.4\% \\
    & Toggle Rate Extraction & 0.8\% \\
    % \multicolumn{2}{|c||}{SOG Construction}      & 3.4               \\
    \hline
    \multirow{3}{*}{ML Feature Generation} & Timing Modeling & 0.3\%               \\
                                & Power Modeling  & 0.1\%               \\
                                & Area Modeling   & 0.1\%               \\
    \hline
    \multicolumn{2}{|c||}{All PPA Prediction Inference Time}      & \textless{}0.001\% \\
    \hline    
    \end{tabular}
    }
    \begin{tablenotes}\scriptsize
        \item $^\star$ The runtime is presented as the proportion of logic synthesis runtime.
    \end{tablenotes}  
    \caption{MasterRTL Runtime Overhead.}  
        \label{tbl:runtime}
\end{table}

\section{Discussions}
\begin{figure}
  \centering
  \includegraphics[height=0.365\linewidth]{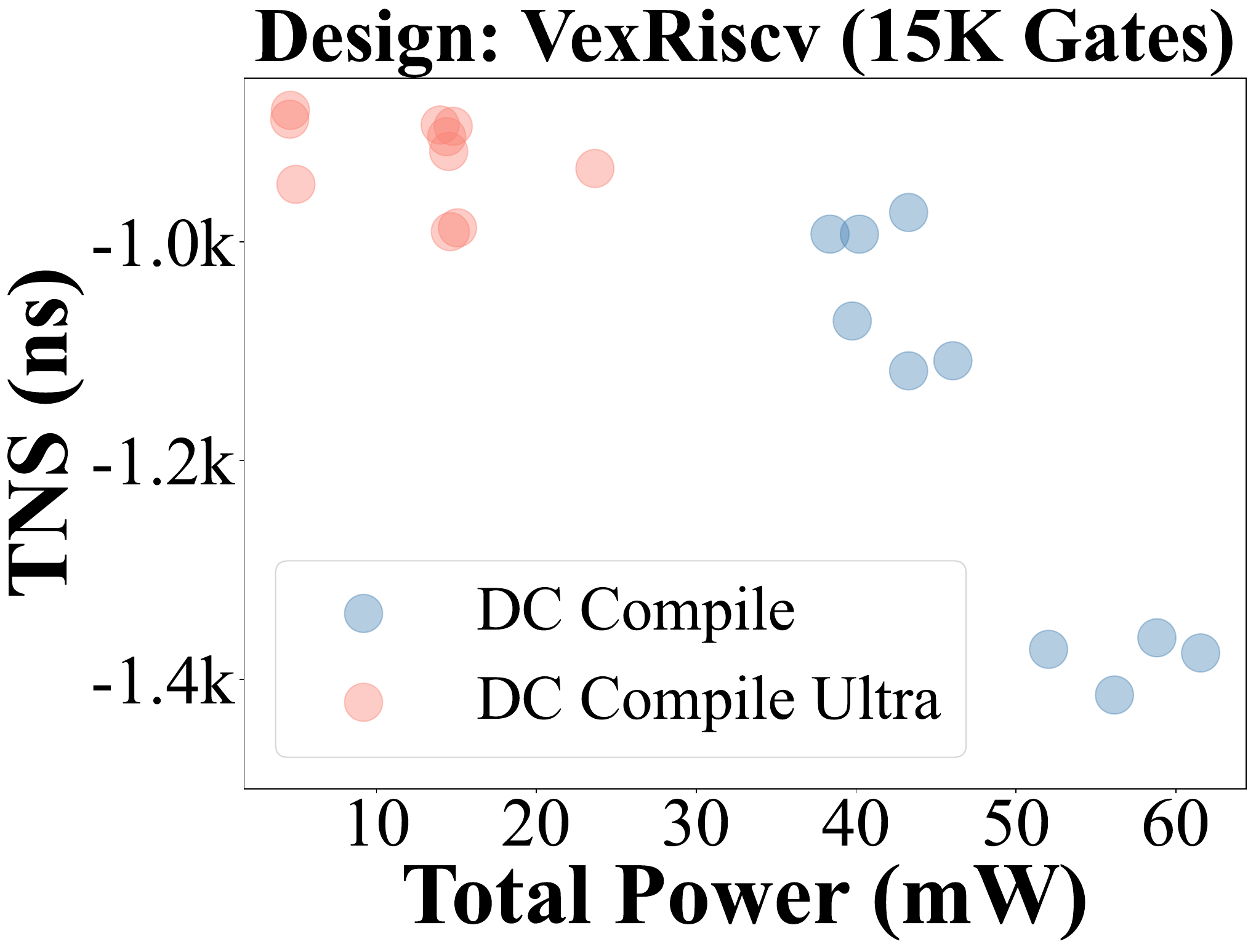}
  \includegraphics[height=0.365\linewidth]{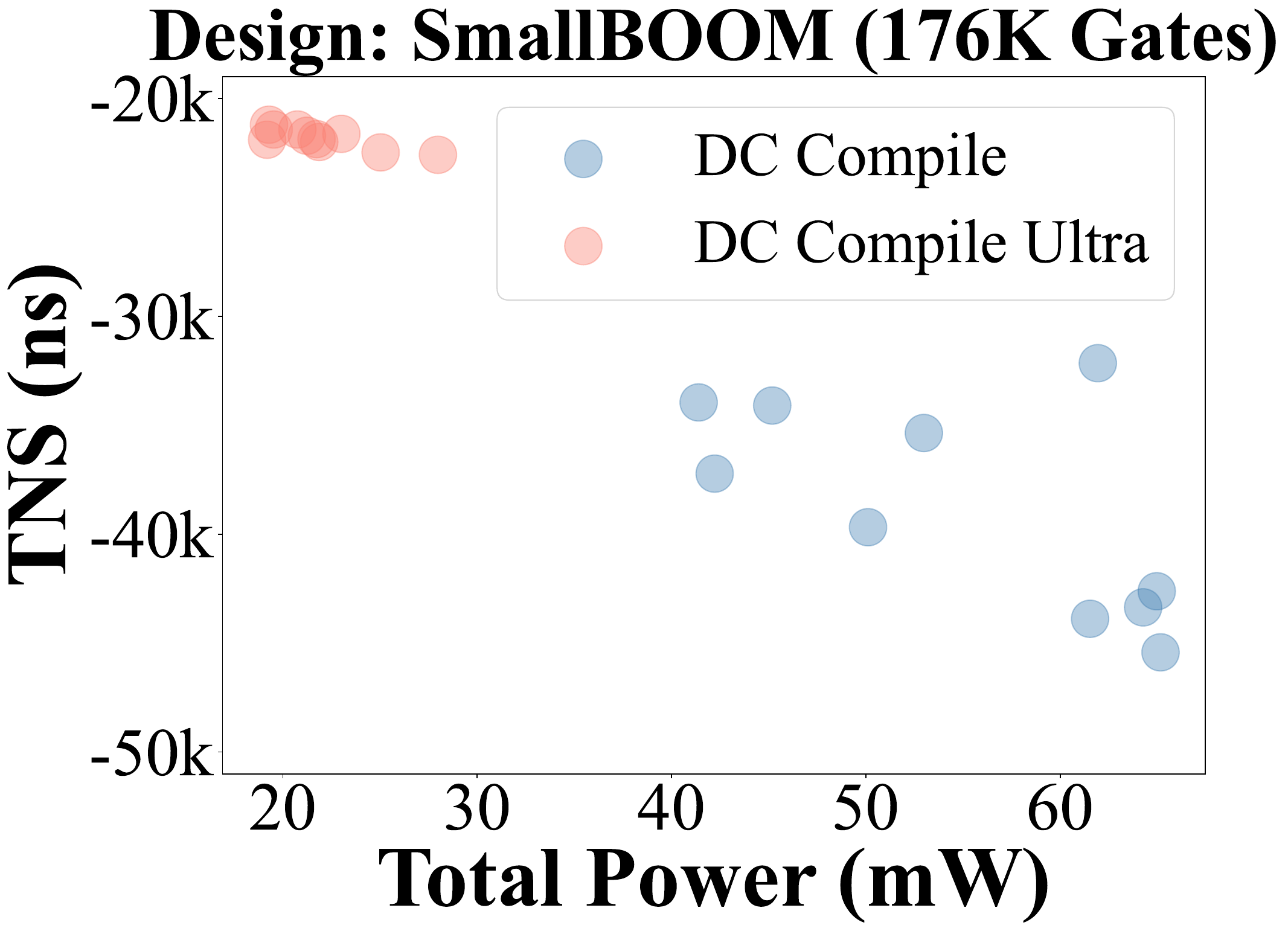}
  \caption{Trade-offs between power and performance caused by different synthesis options. When adopting the most advanced synthesis option (i.e., \texttt{Compile\_Ultra}), there is a best synthesis solution that achieves almost both the best power and performance, without obvious trade-offs. Therefore, this best synthesis solution point is directly adopted to generate the label.}%no obvious trade-offs between design objectives.}
  \label{fig:tradeoff}
  \vspace{-.1in}
\end{figure}

\subsection{MasterRTL Extended to Layout PPA Modeling}

In addition to targeting PPA labels of the post-synthesis netlist $\mathcal{G}$, MasterRTL can be readily extended to predict PPA in the layout stage based on the same RTL designs. Here we first visualize the PPA correlation between the post-synthesis netlist and the post-placement solution from Innovus\textsuperscript{\textregistered}, as shown by green points in Fig.~\ref{fig:layout}. %Initially, we identified the disparities between the ground truth PPA values in the post-synthesis and post-placement stages, represented by the green scattered points. 
These disparities primarily stem from offsets, while maintaining a high correlation. Moreover, the physical-design EDA tools significantly optimize the timing characteristics with the cost of a deterioration in power consumption, meanwhile, the area remains almost unchanged.

Based on this observation, we developed one extra tree-based model to further predict post-placement PPA. Its features are simply from the existing MasterRTL predictions (blue points). Here we perform RTL-stage prediction targeting post-placement, an extremely challenging task. The final predictions, depicted as red scattered points, indicate our framework can achieve similar or even higher accuracy than ground-truth logic synthesis results.

\subsection{The Impact of Trade-offs}
\label{sec:tradeoff}
With the advanced synthesis option from the latest commercial tools, according to our observation, the trade-offs caused by synthesis parameters are not obvious. We compared the two different synthesis options (i.e., \texttt{Compile\_Ultra} and \texttt{Compile}), and the synthesis results of two design examples are shown in Fig.~\ref{fig:tradeoff}. Notably, the PPA variance attained through the utilization of the \texttt{Compile\_Ultra} command is significantly smaller compared to the alternative option, and there is typically a best result in both design objectives, thus the ground-truth label is directly used. The prior work~\cite{sengupta2022good} claiming the necessity of capturing trade-offs actually adopted \texttt{Compile} options.
% In the prior work . 

%for this extremely challenging task. 

%in estimating layout-stage PPA. 

\subsection{Utilization of Generated Pseudo RTL Designs}
\label{sec:grtl}

\begin{figure}
  \centering
  \includegraphics[width=0.8\linewidth]{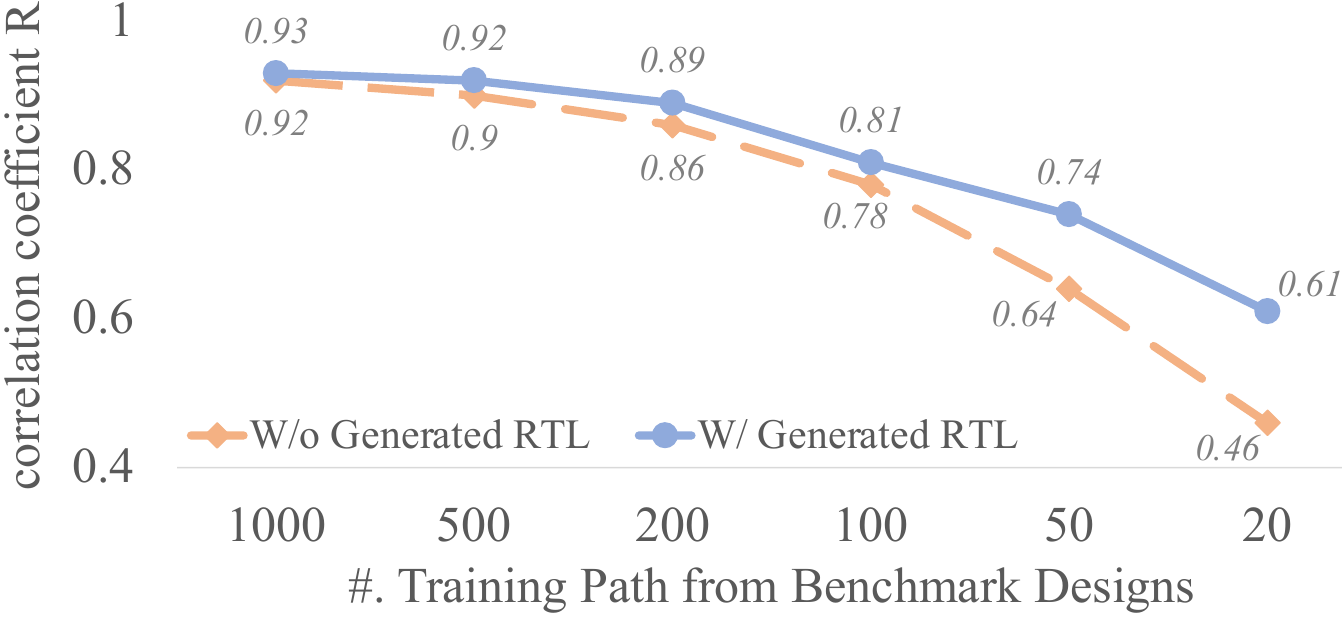}
  \caption{Data augmentation with generated RTL designs in path-level timing modeling. Models with generated RTL perform better, especially when training paths from real RTL designs are limited.}%Reducing training paths leads to decreasing model accuracy. Generated RTL designs mitigate the downward trend.}
  \label{fig:grtl}
  \vspace{-.1in}
\end{figure}
% Introducing pseudo path data from 
% the path dataset in 

% Specifically, the Graph Recurrent Attention Network~\cite{liao2019gran} with only few designs as input, generates new AST-alike graphs, which are further converted back to the Verilog codes. We extracted 220 pseudo paths corresponding to the slack labels from 35 pseudo designs, using the same process as with benchmark designs. 

%are exploited in the training data for the timing path-level models. 

% for data augmentation 
%We demonstrate the impact new generated training data with the path-level model. 
%\footnote{We do not use the complete timing model, since the multiple stages introduce complex impact.}.  
% here to reduce complex impacts from 

Finally, we evaluate the effect of generated RTL designs introduced in Subsection~\ref{sec:dataaug}, when there are insufficient real RTL designs as training data. The experiment evaluates the path slack prediction accuracy of our proposed path-level model, a key component of the timing model. 
Fig.~\ref{fig:grtl} shows its accuracy as the number of training paths from real designs decreases. Without generated paths, the accuracy of the model drops rapidly (orange). But after augmenting the dataset with generated new RTL, there is an improvement (blue), and the gap is more obvious as real-design data further decreases. Generating new RTL designs from scratch obviously has strong practical value, and we will continue to improve the quality of generated RTL in our future works to further enlarge the gap.

\section{Conclusion}

In this paper, we present MasterRTL, a pre-synthesis PPA estimation framework for RTL designs. The proposed method adopts a general RTL representation named simple operator graph (SOG) and customizes multi-stage ML models for WNS, TNS, power, and area separately. Accurate estimations are provided for both the post-synthesis and post-placement stages. Additionally, a data augmentation methodology is demonstrated to address the data availability problem, with plans for future extensions.

\section{Acknowledgement}

This work is partially funded by the Hong Kong Research Grants Council (RGC) ECS Grant 26208723, Guangdong Basic and Applied Basic Research Foundation no. 2022A1515110178, Guangzhou-HKUST(GZ) Joint Funding Scheme no. SL2022A03J01288, Guangzhou Basic Research Project no. SL2022A04J00615, and ACCESS – AI Chip Center for Emerging Smart Systems, sponsored by InnoHK funding, Hong Kong SAR. Also, the authors thank the help from Prof Jiang Hu and Prianka Sengupta at Texas A\&M University.

% \newpage
\bibliographystyle{IEEEtran}
\bibliography{references_1, references_2}
\end{document}